\documentclass[prb,twocolumn,showpacs,superscriptaddress,floatfix, nofootinbib]{revtex4-2}
\usepackage{amsmath,amssymb,amsfonts,mathrsfs, amsbsy}
\usepackage{graphicx}
\usepackage[bookmarks=true, colorlinks=true, linkcolor=blue, urlcolor=blue, citecolor=blue, bookmarks=true, hyperindex=true]{hyperref}
\usepackage[normalem]{ulem}
\usepackage{physics}
\usepackage{comment}
\usepackage{subcaption}
\usepackage[dvipsnames]{xcolor}
\usepackage[utf8]{inputenc}

\newcommand{\be}{\begin{equation}}
\newcommand{\ee}{\end{equation}}

\newcommand{\beq}{\begin{eqnarray}}
\newcommand{\eeq}{\end{eqnarray}}

\def\H1{\widehat{H}_1}

\newcommand{\asp}{\color{blue}}

\begin{document}

\title{Hilbert space geometry and quantum chaos}

\author{Rustem Sharipov}\thanks{These authors contribute equally}
\affiliation{Physics Department, Faculty of Mathematics and Physics, University of Ljubljana,
Ljubljana, Slovenia}

\author{Anastasiia Tiutiakina}\thanks{These authors contribute equally}
\affiliation{Laboratoire de Physique Theorique et Modelisation, CNRS UMR 8089, CY Cergy
Paris Universite, 95302 Cergy-Pontoise Cedex, France}

\author{Alexander Gorsky}
\affiliation{Laboratory of Complex Networks, Center for Neurophysics and Neuromorphic Technologies,
Moscow, Russia}
\author{Vladimir Gritsev}
\affiliation{Institute for Theoretical Physics, University of Amsterdam, Science Park 904, Postbus 94485, 1090
GL Amsterdam, The Netherlands}

\author{Anatoli Polkovnikov}
\affiliation{ Department of Physics, Boston University, 590 Commonwealth Ave., Boston, MA 02215}


\begin{abstract}
The quantum geometric tensor (QGT) characterizes the Hilbert space geometry of the eigenstates of a parameter-dependent Hamiltonian. In recent years, the QGT and related quantities have found extensive theoretical and experimental utility, in particular for quantifying quantum phase transitions both at and out of equilibrium. Here we consider the symmetric part (quantum Riemannian metric) of the QGT for various multi-parametric random matrix Hamiltonians and discuss the possible indication of ergodic or integrable behaviour. We found for a two-dimensional parameter space that, while the ergodic phase corresponds to the smooth manifold, the  integrable limit marks itself as a singular geometry with a conical defect. Our study thus provides more support for the idea that the landscape of the parameter space yields information on the ergodic-nonergodic transition in complex quantum systems, including the intermediate  phase. 

\end{abstract}

\maketitle

\textit{\textbf{Introduction}}. Over the past few decades, there has been significant research into the statistical properties of quantum systems. This has led to the crucial distinction between two types of phases: quantum chaotic and integrable. One method for identifying whether a system is in a chaotic or integrable regime involves the use of the quantum geometric tensor (QGT). The QGT is a tool that quantifies the responds of a system to changes in its parameters, making it a valuable approach for analyzing these distinct phases \cite{Provost_1980,Zanardi_2006,Zanardi_2007}.



The real part of the QGT defines a Riemann metric in the parameter space known as the quantum metric tensor or quantum information metric, while the imaginary part corresponds to the Berry curvature. This metric measures the distances in the space of the eigenfunctions bundled over the parameter space.  The singular behaviour of the Ricci scalar or a change in the topology of the parameter space are associated with quantum phase transitions in the system \cite{Kolodrubetz:2013mla}. The symmetries of this metric can also be used for the classification of quantum phases \cite{Liska2021}. Some recent examples of the identification of critical points via singularities in the quantum metric can be found in Refs.~\cite{peotta2023quantum,alexandrov2023information}.

More recently, the quantum metric tensor has been used as a probe of quantum chaos~\cite{Pandey}. In particular, it was found that  at integrable points for integrability breaking direction of perturbation the metric tensor generically scales exponentially with the system size, similarly to its scaling in ergodic systems that satisfy the eigenstate thermalization hypothesis. There are possible exceptions, though, related to weak integrability breaking perturbations~\cite{Surace_2023,vanovac2024,orlov2023,Pozsgay2024}. In contrast, for integrability preserving perturbations, the scaling of the QGT was found to be polynomial, reflecting the vanishing of spectral weight at low frequencies~\cite{Sels, Kim2023}.  In general, integrable and ergodic/mixing regimes are separated by a KAM-like chaotic but non-thermalizing region~\cite{Pandey,Sels}. A similar intermediate non-ergodic extended (NEE) phase has also been identified in the context of disordered systems. This phase is characterized by a multifractal structure of eigenstates, first observed in the Rosenzweig-Porter (RP) model \cite{kravtsov2015random},
and later confirmed through the replica approach \cite{venturelli2023replica} and the Krylov basis viewpoint \cite{bhattacharjee2024krylov}, see also Refs.~\cite{Bogomolny2018eigenvalue,Tang2022nonergodic,Khaymovich2021dynamical,vonSoosten2017non,altshuler2023random}. In Refs.~\cite{Sugiura_2021,Kim2023} it was further argued that the geometric tensor becomes highly anisotropic near integrable points, and that such integrable points serve as attractors of geodesic flows determined by the metric.

At least three other measures of the "chaos-intermediate-integrability" patterns have to be mentioned. The OTOC correlators 
\cite{larkin1969quasiclassical} related to the operator growth work in some situations, while the approach based on the Krylov complexity
was suggested in \cite{parker2019universal}.  Finally, the dependence of the inverse participation ratio ${\rm IPR}_q(L)$  on the system size $L$ distinguishes the localized and delocalized states and corresponds to the integrable and non-integrable regimes respectively \cite{evers2008anderson}. 
There are some links between the approaches: the measure similar to the fidelity susceptibility in context of localization was proposed in Ref.~\cite{2015}. The QGT can be related to the late-time behaviour of autocorrelator tensor of the operators $\partial_{\lambda_j} H$ where $\lambda_j$ denotes the coordinates in the parameter space~~\cite{lim2024defining}. For this reason there must be some connection between the operator spreading in Krylov space and the QGT.  At the same time there are clear examples
that restrict the validity of the OTOC and operator growth as chaos measures ~\cite{fine2014absence,kukuljan2017weak,pappalardi2020quantum, lim2024defining}.

In this Letter, we investigate the geometry of the perturbed Hilbert space for matrix Hamiltonians. To explore the metric in the neighbourhood of the chaotic point,  we start with a matrix Hamiltonian for a finite-dimensional system and add random matrix perturbations. This analysis is a generalization of~\cite{Berry_2020}, where only the fidelity susceptibility at a single point of the parameter space was considered,  which is not enough to capture the geometry. To examine the metric near an integrable point, we consider diagonal matrices with independent Gaussian distributions of eigenvalues, also known as the random energy model~\cite{Derrida_1980}. In both cases, we perturb the initial system by a pair of random matrices multiplied by parameters, which play the role of the coordinates. We then evaluate the quantum metric in a two-dimensional parameter space in both cases and find a clear-cut indication of the difference in geometries between the integrable and chaotic cases: there is a universal conical singularity emerging at the integrable point, which is related to a long time tail of the autocorrelation function of these random matrices. Our analysis also clearly identifies the presence of an intermediate regime in the behavior of the metric, which is interpreted most naturally as the NEE phase.

\textbf{{\textit{Quantum Geometric Tensor}}}. Let us begin by briefly recalling the definition of the quantum geometric tensor (QGT) and introducing the relevant notations. Consider the eigenspace of some parameter-dependent Hamiltonian $H(\vec \lambda)$. With the energy eigenvalues $E_{n}(\vec \lambda)$ and eigenvectors $\ket{n(\vec \lambda)}  $ depending on the parameters:
\begin{equation}\label{eq:Shr}
    H(\vec \lambda) \ket{n(\vec \lambda)}=E_n(\vec \lambda) \ket{n(\vec \lambda)},
    \end{equation}
where the parameter $\vec \lambda=\{\lambda_1,\lambda_2,\dots\}$ is generally multi-component.  In the space of eigenvectors one can define the distance between nearby states with infinitesimally different parameters in the following way~\cite{Provost_1980}
\begin{equation}
ds^2 \equiv 1-\left| \braket{n(\vec \lambda)}{n(\vec \lambda+d \vec \lambda)}\right|^2 .
\end{equation}
Then the quantum geometric tensor for the $n$-th eigenstate is defined as the leading contribution in $d\vec\lambda$ expansion of the distance defined above:
\begin{multline}
ds^2 \equiv g^{(n)}_{\alpha \beta} d\lambda_\alpha d\lambda_\beta+\mathcal{O}(|\vec{d \lambda|}^3)=\\=
\braket{\partial_{\alpha} n}{\partial_{\beta} n}-\braket{\partial_{\alpha} n }{n}\braket{n}{\partial_{\beta} n}+\mathcal{O}(|\vec{d \lambda|}^3), 
\label{QGT}
\end{multline}
where  $\partial_\alpha \equiv \partial_{\lambda_\alpha}$. Notice, that the quantum geometric tensor is invariant under arbitrary phase transformations of the eigenfunctions $\ket{n(\lambda)}\rightarrow e^{i \phi_n(\lambda)} \ket{n(\lambda)}$, that is expected from the gauge invariance of the QGT. In the absence of degeneracies, it is straightforward to rewrite (\ref{QGT}) in the following form (see Appendix \ref{App:Ham}):
\begin{equation}\label{Geom}
g_{\alpha \beta}^{(n)}=\sum\limits_{m \neq n}\frac{\left< n |\partial_{\alpha }H |m\right>\left< m| \partial_\beta H|n \right>}{(E_n-E_m)^2},
\end{equation}
which we will use in the next sections. 

The real part of quantum geometric tensor is the quantum metric tensor, while the imaginary part is the Berry curvature. In this work we will focus on the {\it averaged geometric tensor} (or the Hilbert–Schmidt norm of the adiabatic gauge potential~\cite{Pandey}):
\begin{eqnarray}\label{AGP}
G_{\alpha \beta} =\frac{1}{N} \sum\limits_{n=1}^{N}g_{\alpha \beta}^{(n)},
\end{eqnarray}
where $N$ is Hilbert-space dimension, e.g. $N=2^L$ for qubit spin chains. It is straightforward to verify that the imaginary, antisymmetric part of this averaged geometric tensor is zero when the parameter-dependent Hamiltonian is Hermitian. Additionally,  we note that the Hilbert space average is equivalent to the average with respect to the infinite temperature density matrix.

{\textbf{\textit{Geometry of a Random Matrix Model}}}. Following Berry and Shukla \cite{Berry_2020}, we consider a two-parameter family of  $N\times N$ random Hamiltonians:
\begin{equation}\label{ham}
    H = H_0 + x H_x + y H_y.
\end{equation}
We focus on the case when random matrices $H_0$, $H_x$, and $H_y$ are independently drawn from the Gaussian Unitary Ensemble (GUE) with the distribution
\be
\rho(H_{a})=e^{-
 \frac{N}{2} Tr \left( H_{a}^2 \right)}, \quad a=0,x,y.
\ee

In this section, we calculate the QGT, for this purely random model. We treat the QGT as an induced metric on a two-dimensional surface embedded in the three-dimensional Euclidean space. First, notice that it is convenient to move to the polar coordinates, since our system has no preferred direction on average. That is, we redefine:
\begin{equation}\label{hamr}
    H = H_0 + r \cos{\phi}~ H_x + r \sin{\phi}~H_y .
\end{equation}

 It is clear that the metric tensor averaged over the random matrices $H_0, H_x, H_y$, must be independent of the angle $\phi$. In the course of straightforward calculations (see Appendix \ref{calcul}), the components of the averaged QGT are given by:
\begin{equation}\label{chaosmetr}
    \overline{G}_{rr}(r)=\frac{N-1}{2(r^2+1)^2},~~~\overline{G}_{\phi \phi}(r)=r^2 \frac{N-1}{2(r^2+1)}, 
\end{equation}
 and $\overline{G}_{r \phi}=0$.

\begin{figure}[h!]
\centering
\centering{\includegraphics[scale=0.16,trim={15cm 8cm 15cm 19cm},clip]{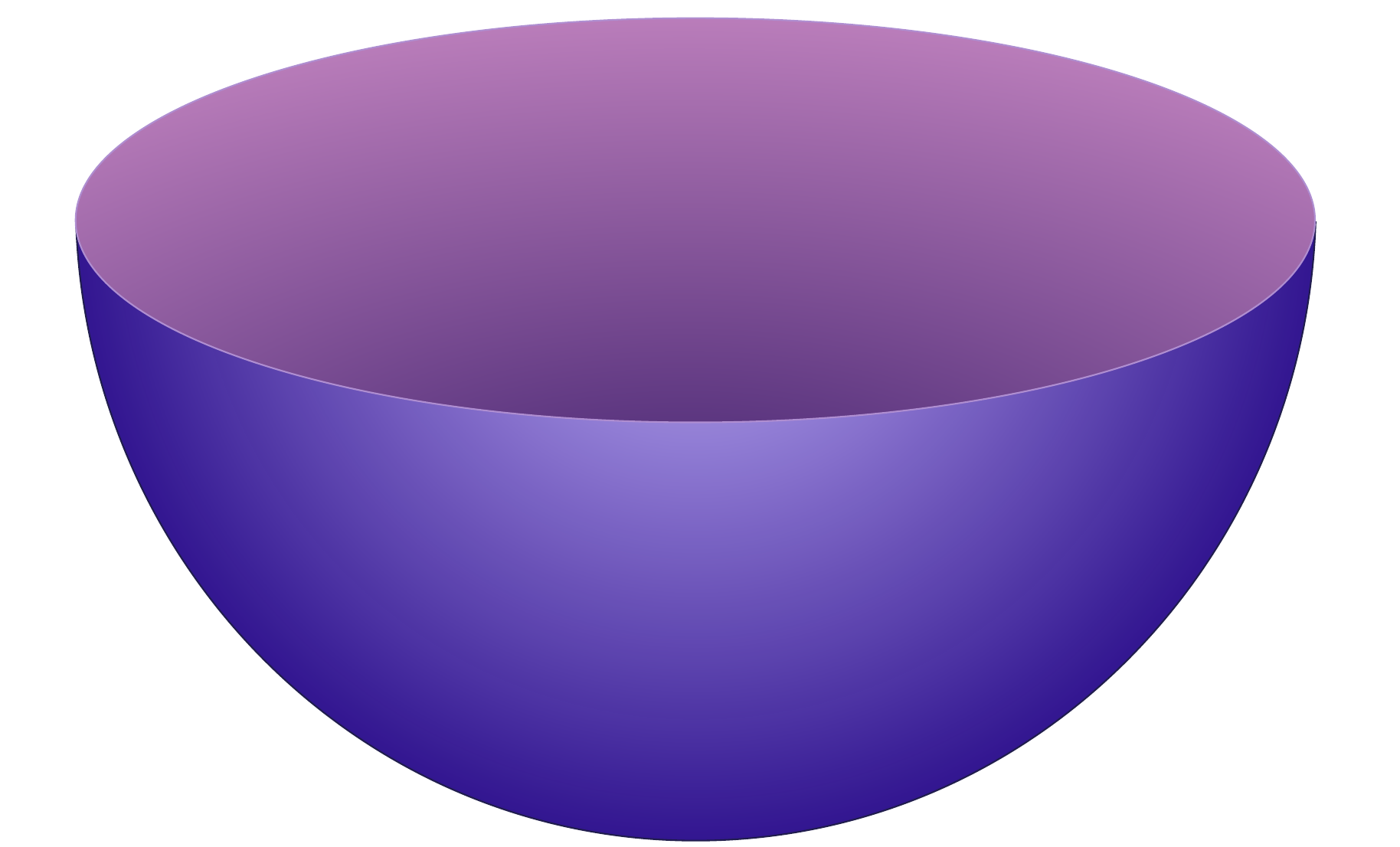}}
\caption{Isometric manifold $\mathcal{M}_{ch}$}
\label{Mchaos}
\end{figure}

\textit{\textbf{Topology by embedding}}. Finally, the two-dimensional first fundamental form can be written as
\begin{equation}
ds^2= \overline{G}_{rr}(r) dr^2+\overline{G}_{\phi \phi}(r) d\phi^2.
\end{equation}
We analyse the topological characteristics of the corresponding manifold by constructing an isometric surface and plotting its shape. In order to do this we consider three-dimensional Euclidean space and two-dimensional submanifold $\mathcal{M}_{ch}$ embedded within it, with the metric set by the standard relation:
\begin{equation}\label{metrind}
dZ^2+dR^2+R^2 d\phi^2=\overline{G}_{rr}(r) dr^2+\overline{G}_{\phi \phi}(r) d\phi^2.
\end{equation}
Since the metric components are independent of $\phi$, the submanifold $\mathcal{M}_{ch}$ can be parametrized as follows: 
\begin{equation}
\mathcal{M}_{ch}:~~Z=Z(r),~R=R(r).\nonumber
\end{equation} Then by using Eq. (\ref{metrind}) we obtain the corresponding system of equations, which reads as
\begin{subequations}\label{Emb}
\begin{equation}
\left( \frac{dZ}{dr}\right)^2+\left( \frac{dR}{dr}\right)^2=\overline{G}_{rr}(r),
\end{equation}
\begin{equation}
R(r)^2=\overline{G}_{\phi \phi}(r).
\end{equation}\end{subequations}
This system of equations can be easily solved for the metric tensor we derived:
\begin{equation}
\begin{split}
&R(r)=R_0 \frac{r}{\sqrt{r^2+1}},~~ Z(r)=R_0\left( 1-\frac{1}{\sqrt{r^2+1}}\right),
\end{split}
\label{eq:metric_rmt}
\end{equation}
where $R_0=\sqrt{\frac{N-1}{2}}$. We fixed the integration constant by assuming $Z(0)=0$. This system defines a two-dimensional surface parametrised by $r$. By excluding $r$ from these equations, we find that $Z$ and $R$ satisfy the following constraint relation:
\begin{equation}
\left(Z-R_0\right)^2+R^2=R_0^2,
\end{equation}
where $0 \leq R\leq R_0$ and $0\leq Z \leq  R_0$ . This represents the equation of a lower hemisphere centered at $Z=R_0$ with radius $ R_0$. In Fig. \ref{Mchaos} we illustrate the shape of this isometric manifold.

This result can be anticipated from heuristic reasoning: by adding a random matrix deformation $x H_x+y H_y$ to an existing random matrix $H_0$ we generate another random matrix equivalent to $H_0$ apart from rescaling of the variance $\sigma^2$. The geometry of the manifold cannot depend on the overall scale, therefore locally the shape should be both $r$ and $\phi$ - independent. Hence, $\mathcal M_{ch}$ is locally equivalent to the surface of a sphere, where all points are also equivalent.

\textbf{\textit{Geometry of Integrability Breaking}}. Now, let us consider a different Hamiltonian, where the unperturbed Hamiltonian $H_0$ is replaced by a diagonal matrix $\Lambda_0$ with independent random entries while the perturbation is still a random matrix. Such a diagonal matrix exhibits Poisson level statistics and can therefore be regarded as a representation of an integrable model.. The Hamiltonian $H$ now takes the form
\begin{equation}
\label{eq:H_def}
H=\Lambda_0+x H_x+y H_y,
\end{equation}
where $\Lambda_0={\rm diag}(\lambda_i)$ with $\rho(\lambda_i)=e^{-\frac{1}{2} \lambda_i^2}$, and  $H_x$ and $H_y$ are drawn from GUE as before. This setup is comparable to the  RP model~\cite{kravtsov2015random} with the addition that we also allow deformations along the $\phi$-direction, which obviously keeps the spectrum invariant but has a non-trivial metric associated with eigenstate deformations. For $N=2$ the model is similar to the one considered in \cite{motamarri2022localization}, where the angular parameter is also present. It is intuitively clear that the radial and angular directions should no longer be equivalent near the integrable point, and thus, the spherical geometry found in the previous section must be distorted near $r=0$.

Let us first consider $N=2$, where the entire analysis can be carried out analytically. We find that the quantum metric tensor in polar coordinates is given by:
\begin{subequations}
\begin{equation}
\label{eq:G_rr_int}
\overline{G}_{\phi \phi}=r \frac{1}{2~\sqrt{2} } \arctan\left( \frac{\sqrt{2}}{r}\right),
\end{equation}
\begin{equation}
\label{eq:G_ff_int}
\overline{G}_{r r}=\frac{1}{4} \left(\frac{\arccot\left(\frac{r}{\sqrt{2}}\right)}{\sqrt{2}r}-\frac{1}{2+r^2} \right),
\end{equation}
\end{subequations}
and $\overline{G}_{r \phi}=0$. The fact that the metric tensor is   $\phi$  independent and that its mixed component of it is zero again allows us  to perform the embedding in pseudo-Euclidean space (see Appendix \ref{App:metrn2}).

\begin{figure} [h!]
\centering
\centering{\includegraphics[scale=0.16,trim={15cm 13cm 17cm 10cm},clip]{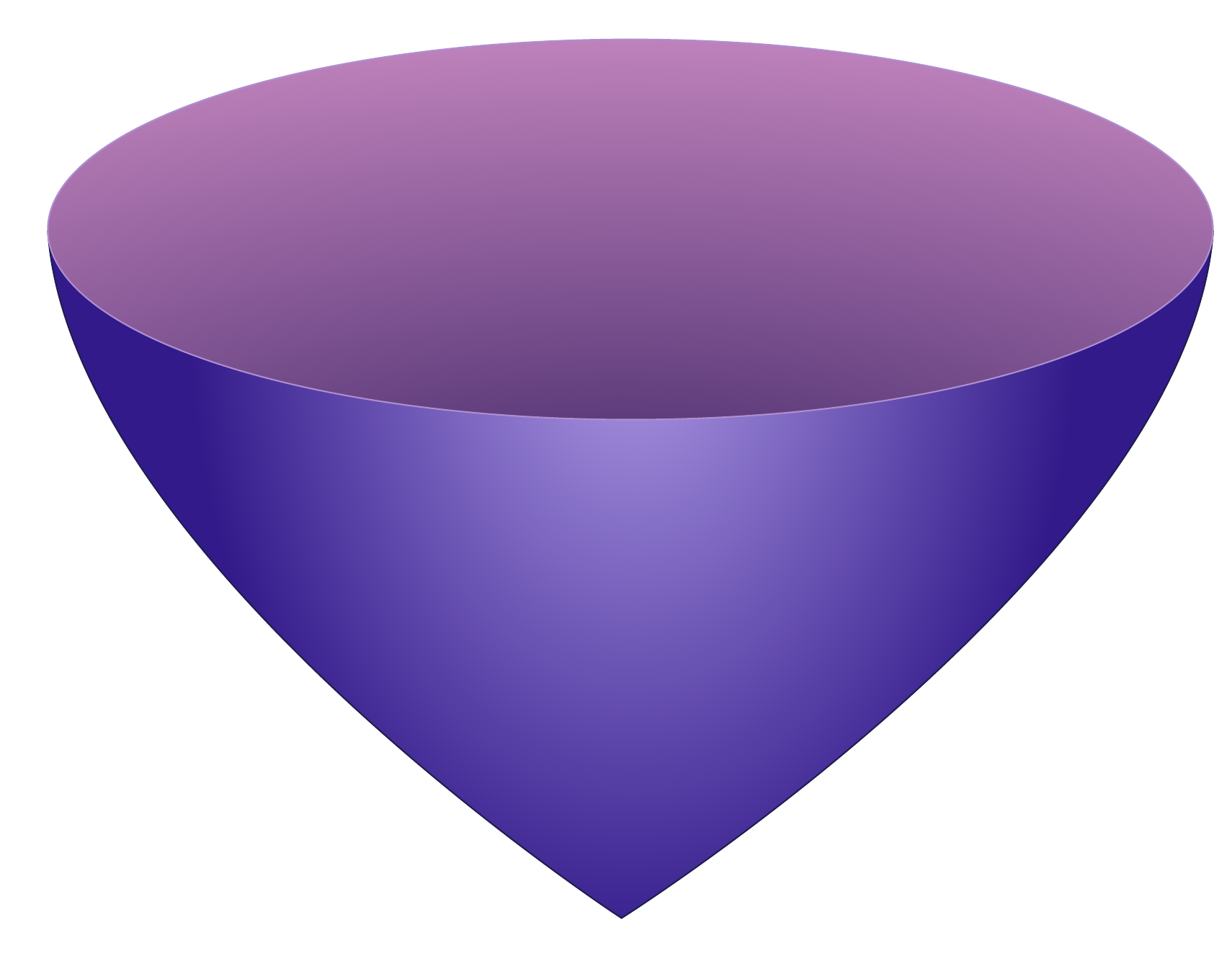}}
\caption{Isometric manifold $\mathcal{M}_{int}$}
 \label{cone}
\end{figure}

Near the origin, i.e.. at $r\ll 1$, we can Taylor expand the metric and obtain the following expression for the first fundamental form
\begin{equation}
    \label{con2d}
    ds^2 \approx \frac{\pi}{8 \sqrt{2}}\left( \frac{dr^2}{r}+2r d\phi^2\right)=
    \frac{\pi}{2 \sqrt{2}}\left( d\rho^2+\frac{1}{2}\rho^2 d\phi^2\right), 
\end{equation}
where $\rho=\sqrt{r}$. Embedding this metric, as before, into Euclidean space, we find
\begin{equation}
    R(\rho)\approx Z(\rho)\approx \sqrt{\frac{\pi}{4\sqrt{2}}}\rho
\end{equation}
This surface represents a cone with an angle $\pi/2$ (see Appendix \ref{Conical}), we illustrate the shape of isometric manifold in Fig.\ref{cone}. In the limit of $r\gg 1$, the components of the metric tensor~\eqref{eq:G_rr_int} and~\eqref{eq:G_ff_int} approach those of the random model~\eqref{chaosmetr}, and the shape crossovers to the hemisphere. The presence of the conical singularity indicates a change in the geometric properties of the quantum manifold. In particular, the quantum eigenstates become parametrically more sensitive to small perturbations in the $\phi$-direction than to the radial (random) direction close to the integrable point. 
\begin{figure}[h!]
  \centering
  \begin{subfigure}[t]{\linewidth}
  \centering
  \includegraphics[width=1\linewidth]{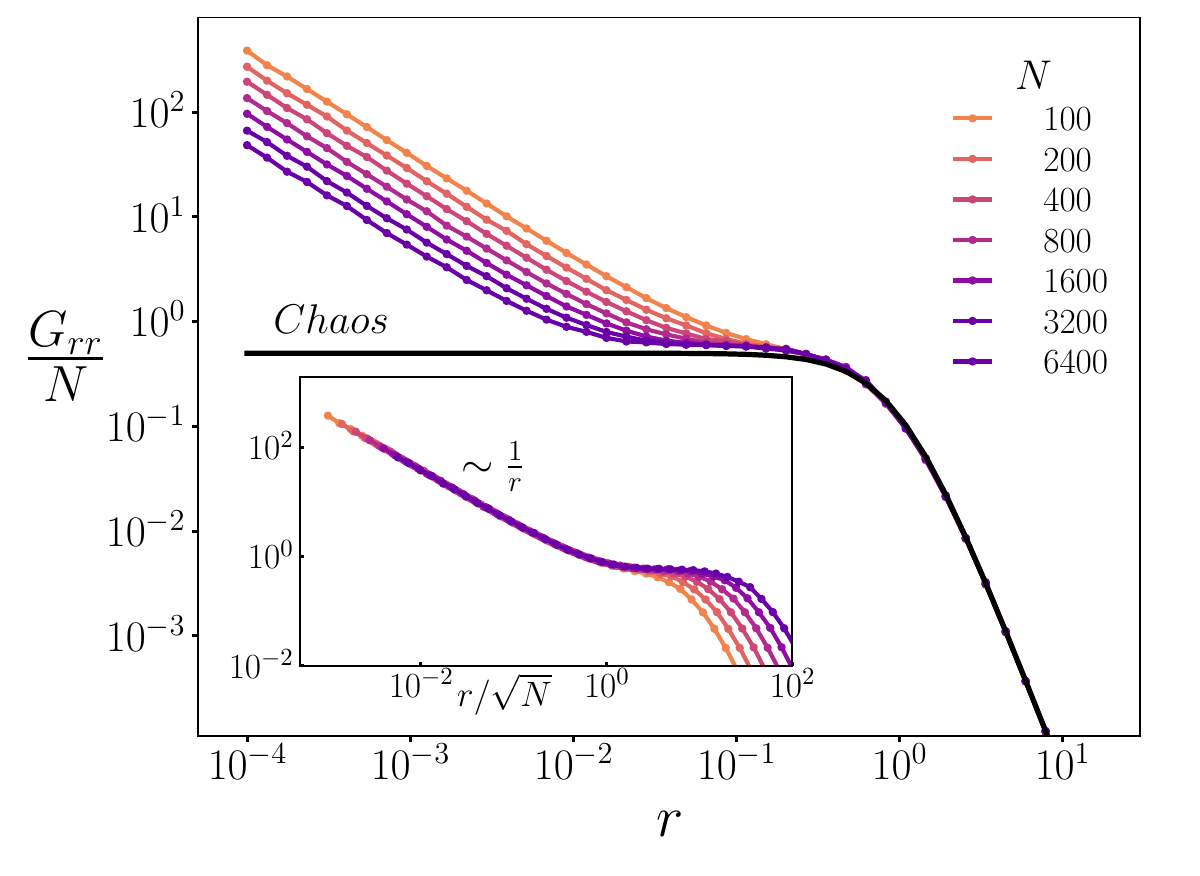}
  \end{subfigure}
  \begin{subfigure}[t]{\linewidth}
    \centering
    \includegraphics[width=1\linewidth]{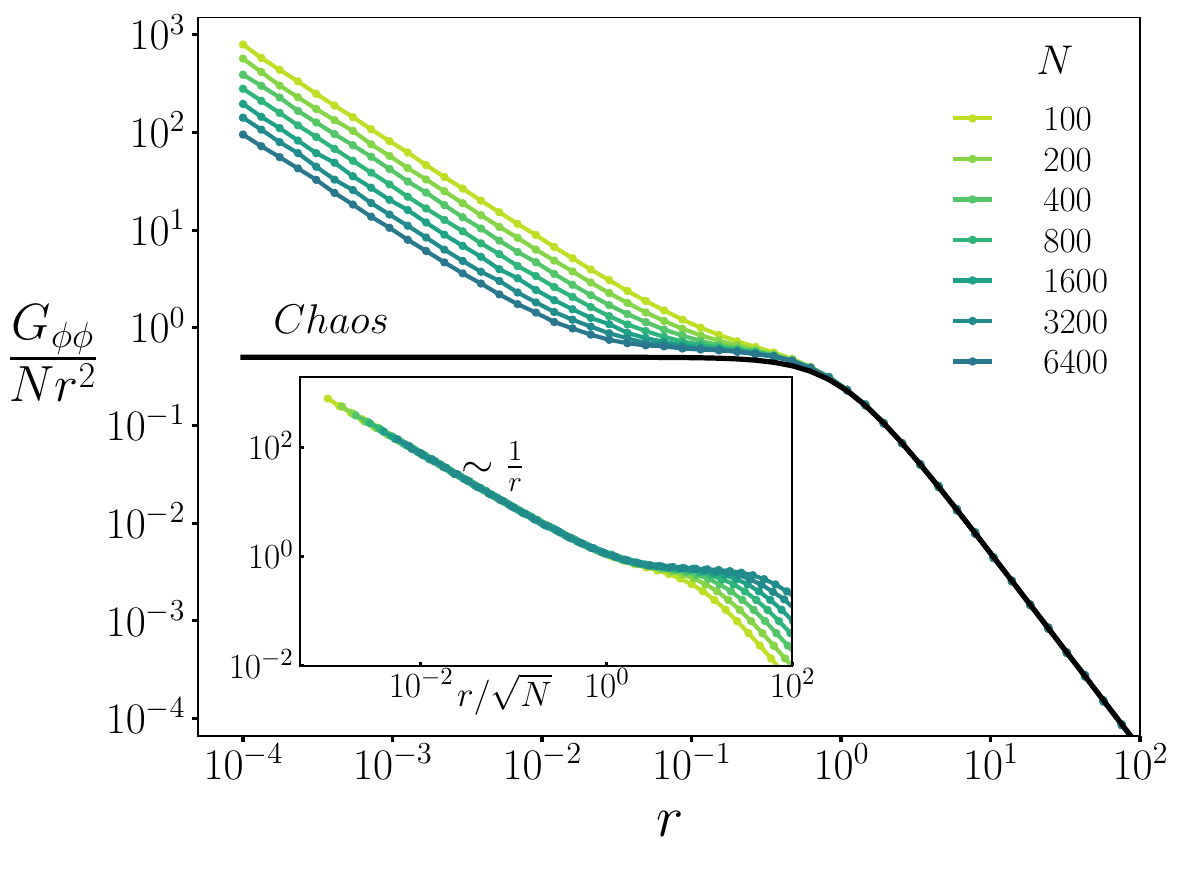}
  \end{subfigure}
  \caption{$G_{rr}$ and $G_{\phi \phi}$ components of the QGT as a function of radial parameter $r$. The inset on the plots represents metric components as function of rescaled parameter $r/\sqrt{N}$. The black line corresponds to the fully chaotic case results given by Eq.(\ref{chaosmetr}).}
  \label{figmetr}
\end{figure}
Having identified the conical geometry at the origin for $N=2$, let us turn to numerical simulations, increasing the system size. The corresponding behaviour of the metric components near the origin  is presented in Fig \ref{figmetr}. One can identify three different regimes, consistent with Ref.~\cite{skvortsov2022sensitivity}.  For $r\ll r^\ast=1/\sqrt{N}$, there is an evident $1/r$ behaviour of the metric, similar to the $N=2$ case. Here
\begin{equation}
    ds^2 \sim \sqrt{N}\left(  \frac{dr^2}{r}+2r d\phi^2 \right),
\end{equation}
which, apart from the overall prefactor, agrees with Eq.~\eqref{con2d}. This regime corresponds to the conical geometry. We note that $r^\ast$ is the localization transition point, so physically, the conical geometry describes the localized phase. For large $r\gg 1$, which corresponds to the ergodic phase, $G_{rr}$ and $G_{\phi \phi}$ approach the corresponding values of the RMT model~\eqref{chaosmetr}. This regime obviously represents spherical geometry. 
\begin{figure}[h!]
  \centering
  \begin{subfigure}[t]{\linewidth}
  \centering
  \includegraphics[width=1\linewidth]{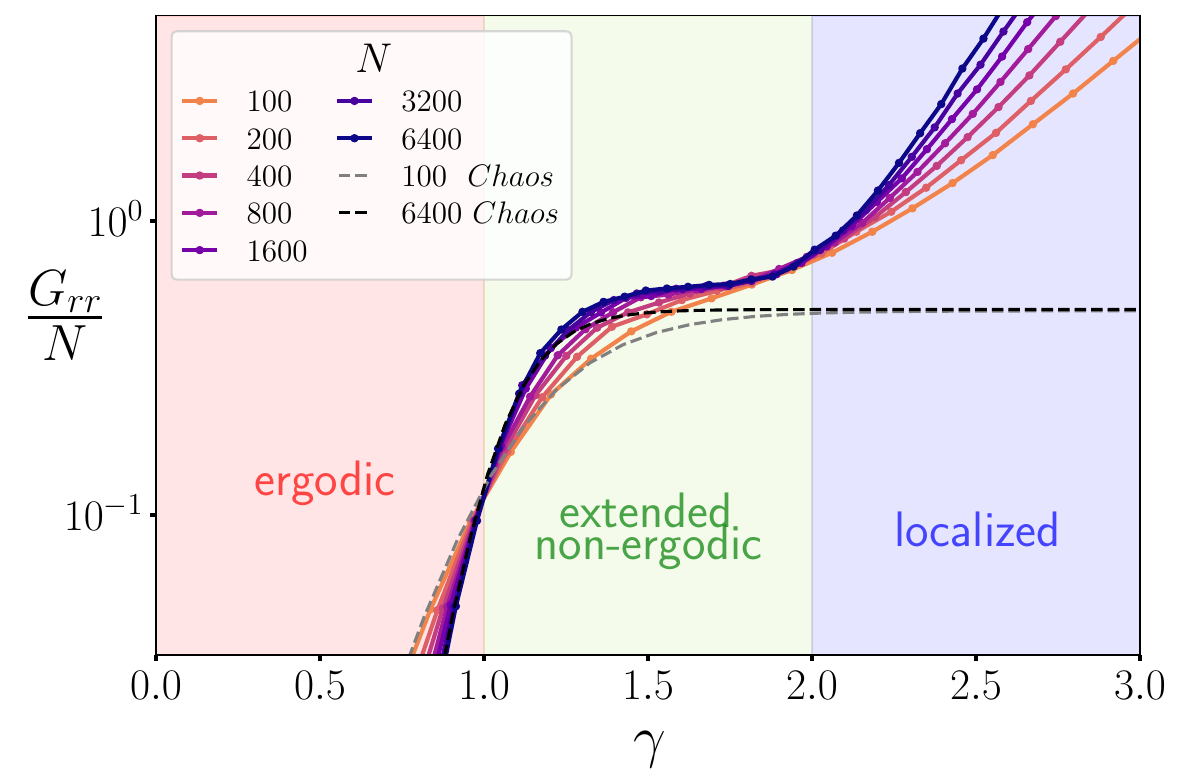}
  \end{subfigure}
  \begin{subfigure}[t]{\linewidth}
    \centering
    \includegraphics[width=1\linewidth]{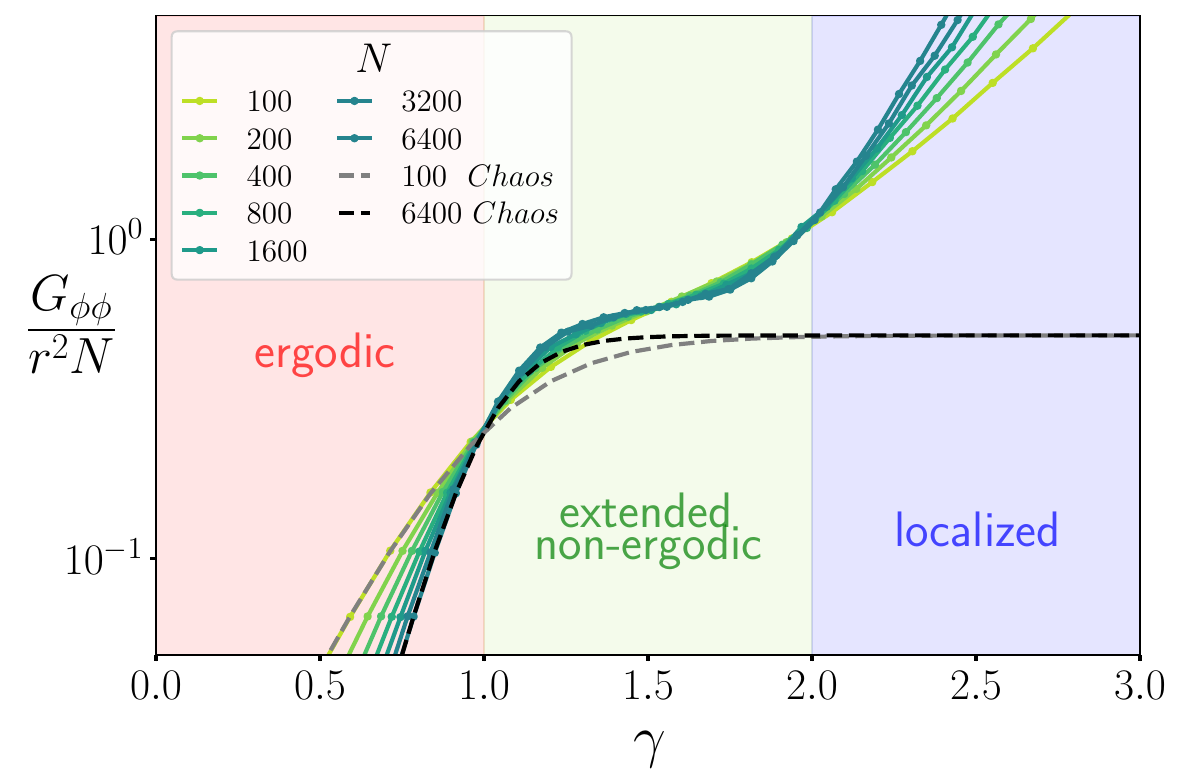}
  \end{subfigure}
  \caption{Components of QGT as a function of scaling parameter}
  \label{large2}
\end{figure}

Finally, there is an intermediate regime at $r^\ast\ll r \ll 1$ representing delocalized non-ergodic phase~\cite{skvortsov2022sensitivity}, where $G_{rr}$ and $G_{\phi\phi}/r^2$ saturate at constants scaling as $N$, but with the prefactors somewhat different from the RMT model. We found that in this intermediate regime, the geometry is still spherical or very close to spherical. To better identify the intermediate regime, we reparametrise $r=N^{(1-\gamma)/2}$ such that the localized, ergodic, and the intermediate phases correspond to $\gamma>2$, $\gamma<1$, $1<\gamma<2$, respectively. Fig. \ref{large2} clearly indicates that there are three different scaling regimes for the metric corresponding to these three phases.

It is worth noting the apparent analogy between the conical singularity near the integrable point and that (with the same conical angle) found previously in Ref.~\cite{Kolodrubetz:2013mla} for the ground state metric close to the isotropic quantum critical point in the ferromagnetic regime of the anisotropic $XY$ spin chain. Our study shows an apparent similarity between the integrable and quantum critical points, with both characterized by the critical slowing down of dynamics and hence by divergent metric, as well as by a strong anisotropy along the radial and azimuthal directions for perturbations. In both situations, we observe condensation of the level crossings, near the origin (i.e. near the critical or integrable point), which are equivalent to magnetically charged defects producing vortex Berry connections in two-dimensional parameter space or the monopole Berry connection in three-dimensional parameter space.

For random matrix perturbations considered in this work, the matrix elements entering Eq.~\eqref{Geom} are random, and hence the scaling of the geometric tensor is entirely determined by the spectral properties of the Hamiltonian. In this 
case, the diagonal components of the geometric tensor become equivalent to the spectral complexity $C(t)$, defined in Sec. 7 of Ref.~\cite{iliesiu2021}. The inverse of the time $t$ used to regularize the spectral complexity is equivalent to the energy cutoff $\mu$ introduced in Refs.~\cite{Pandey,Sugiura_2021}. Physically the regularized geometric tensor determines distance between energy shells of width $\mu$. Then, $t$ or $1/\mu$ represent the minimal time required to measure this regularized metric. From these considerations we conclude that the $1/r$ divergence of the geometric tensor leading to the conical singularity (see Fig.~\ref{large2}) originates from the loss of the level repulsion and hence corresponds to $\mu$ of the order of the level spacing: $\mu\sim 1/N$ (see Appendix~\ref{App:mu} for details).

\textit{\textbf{Conclusions and Outlook.}}
In this Letter, we discussed the geometry of the two-dimensional parameter spaces for matrix Hamiltonians. Combining the analytical and numerical tools, we found that the associated geometry is smooth and locally equivalent to the surface of the sphere when the perturbation is large and the model is effectively in the ergodic regime. At the same time, near the integrable point, the metric develops a conical defect both for the Hilbert space dimension $N=2$ and large $N$ matrices. We also observed three different regimes of the metric behaviour, corresponding to the localized, ergodic, and intermediate  phases. The observed conical singularity is reminiscent of the one observed for the ground state near a continuous quantum phase transition. This similarity highlights the close analogy between the integrable and quantum critical points, both characterized by the emergence of divergent time scales and singular geometry.

 It would be interesting to analyze in more detail the geometry by adding "integrable" perturbations commuting with the unperturbed Hamiltonian. There, the metric is expected to diverge even more strong due to an additional enhancement of the matrix elements of the perturbation near the integrable point~\cite{Sels,skvortsov2022sensitivity}. We leave the corresponding analysis of the associated geometry for future work. Another interesting direction concerns the  investigation of the metric for the perturbed matrix Russian Doll Model with TRS breaking, which also hosts
the  NEE  phase \cite{motamarri2022localization}. Finally, we note that the adiabatic gauge potential and hence the metric can be directly computed from 
the Lanczos coefficients defining the Krylov basis~\cite{bhattacharjee2023,Takahashi_2024}. For this reason, it can be also very interesting to study geometry for the
matrix Hamiltonians enjoying  the exact results \cite{dumitriu2002matrix,balasubramanian2023tridiagonalizing}.  The large $N$ matrix Hamiltonians
are also studied in the context of low-dimensional holography as Hamiltonians in the boundary 1d model, see for instance \cite{saad2019jt}. Hence, it would be interesting to recognize the meaning of our findings in the dual 2d gravity.

\begin{acknowledgments}
We thank C.Jonay, T.Prosen, S.Grozdanov, L.Vidmar, P.Łydżba and D.Kurlov for useful discussions. R.S. acknowledges support from ERC Advanced grant No. 101096208– QUEST, and Research Programme P1-0402 of Slovenian Research and Innovation Agency (ARIS).
A.T. is founded by the ERC Starting Grant 101042293 (HEPIQ)
\end{acknowledgments}

\bibliography{references}

\begin{thebibliography}{42}%
\makeatletter
\providecommand \@ifxundefined [1]{%
 \@ifx{#1\undefined}
}%
\providecommand \@ifnum [1]{%
 \ifnum #1\expandafter \@firstoftwo
 \else \expandafter \@secondoftwo
 \fi
}%
\providecommand \@ifx [1]{%
 \ifx #1\expandafter \@firstoftwo
 \else \expandafter \@secondoftwo
 \fi
}%
\providecommand \natexlab [1]{#1}%
\providecommand \enquote  [1]{``#1''}%
\providecommand \bibnamefont  [1]{#1}%
\providecommand \bibfnamefont [1]{#1}%
\providecommand \citenamefont [1]{#1}%
\providecommand \href@noop [0]{\@secondoftwo}%
\providecommand \href [0]{\begingroup \@sanitize@url \@href}%
\providecommand \@href[1]{\@@startlink{#1}\@@href}%
\providecommand \@@href[1]{\endgroup#1\@@endlink}%
\providecommand \@sanitize@url [0]{\catcode `\\12\catcode `\$12\catcode `\&12\catcode `\#12\catcode `\^12\catcode `\_12\catcode `\%12\relax}%
\providecommand \@@startlink[1]{}%
\providecommand \@@endlink[0]{}%
\providecommand \url  [0]{\begingroup\@sanitize@url \@url }%
\providecommand \@url [1]{\endgroup\@href {#1}{\urlprefix }}%
\providecommand \urlprefix  [0]{URL }%
\providecommand \Eprint [0]{\href }%
\providecommand \doibase [0]{https://doi.org/}%
\providecommand \selectlanguage [0]{\@gobble}%
\providecommand \bibinfo  [0]{\@secondoftwo}%
\providecommand \bibfield  [0]{\@secondoftwo}%
\providecommand \translation [1]{[#1]}%
\providecommand \BibitemOpen [0]{}%
\providecommand \bibitemStop [0]{}%
\providecommand \bibitemNoStop [0]{.\EOS\space}%
\providecommand \EOS [0]{\spacefactor3000\relax}%
\providecommand \BibitemShut  [1]{\csname bibitem#1\endcsname}%
\let\auto@bib@innerbib\@empty
\bibitem [{\citenamefont {Provost}\ and\ \citenamefont {Vallee}(1980)}]{Provost_1980}%
  \BibitemOpen
  \bibfield  {author} {\bibinfo {author} {\bibfnamefont {J.~P.}\ \bibnamefont {Provost}}\ and\ \bibinfo {author} {\bibfnamefont {G.}~\bibnamefont {Vallee}},\ }\bibfield  {title} {\bibinfo {title} {Riemannian structure on manifolds of quantum states},\ }\href {https://doi.org/10.1007/bf02193559} {\bibfield  {journal} {\bibinfo  {journal} {Communications in Mathematical Physics}\ }\textbf {\bibinfo {volume} {76}},\ \bibinfo {pages} {289} (\bibinfo {year} {1980})}\BibitemShut {NoStop}%
\bibitem [{\citenamefont {Zanardi}\ and\ \citenamefont {Paunkovi\ifmmode~\acute{c}\else \'{c}\fi{}}(2006)}]{Zanardi_2006}%
  \BibitemOpen
  \bibfield  {author} {\bibinfo {author} {\bibfnamefont {P.}~\bibnamefont {Zanardi}}\ and\ \bibinfo {author} {\bibfnamefont {N.}~\bibnamefont {Paunkovi\ifmmode~\acute{c}\else \'{c}\fi{}}},\ }\bibfield  {title} {\bibinfo {title} {Ground state overlap and quantum phase transitions},\ }\href {https://doi.org/10.1103/PhysRevE.74.031123} {\bibfield  {journal} {\bibinfo  {journal} {Phys. Rev. E}\ }\textbf {\bibinfo {volume} {74}},\ \bibinfo {pages} {031123} (\bibinfo {year} {2006})}\BibitemShut {NoStop}%
\bibitem [{\citenamefont {Zanardi}\ \emph {et~al.}(2007)\citenamefont {Zanardi}, \citenamefont {Giorda},\ and\ \citenamefont {Cozzini}}]{Zanardi_2007}%
  \BibitemOpen
  \bibfield  {author} {\bibinfo {author} {\bibfnamefont {P.}~\bibnamefont {Zanardi}}, \bibinfo {author} {\bibfnamefont {P.}~\bibnamefont {Giorda}},\ and\ \bibinfo {author} {\bibfnamefont {M.}~\bibnamefont {Cozzini}},\ }\bibfield  {title} {\bibinfo {title} {Information-theoretic differential geometry of quantum phase transitions},\ }\href {https://doi.org/10.1103/PhysRevLett.99.100603} {\bibfield  {journal} {\bibinfo  {journal} {Phys. Rev. Lett.}\ }\textbf {\bibinfo {volume} {99}},\ \bibinfo {pages} {100603} (\bibinfo {year} {2007})}\BibitemShut {NoStop}%
\bibitem [{\citenamefont {Kolodrubetz}\ \emph {et~al.}(2013)\citenamefont {Kolodrubetz}, \citenamefont {Gritsev},\ and\ \citenamefont {Polkovnikov}}]{Kolodrubetz:2013mla}%
  \BibitemOpen
  \bibfield  {author} {\bibinfo {author} {\bibfnamefont {M.}~\bibnamefont {Kolodrubetz}}, \bibinfo {author} {\bibfnamefont {V.}~\bibnamefont {Gritsev}},\ and\ \bibinfo {author} {\bibfnamefont {A.}~\bibnamefont {Polkovnikov}},\ }\bibfield  {title} {\bibinfo {title} {{Classifying and measuring geometry of a quantum ground state manifold}},\ }\href {https://doi.org/10.1103/PhysRevB.88.064304} {\bibfield  {journal} {\bibinfo  {journal} {Phys. Rev. B}\ }\textbf {\bibinfo {volume} {88}},\ \bibinfo {pages} {064304} (\bibinfo {year} {2013})},\ \Eprint {https://arxiv.org/abs/1305.0568} {arXiv:1305.0568 [cond-mat.stat-mech]} \BibitemShut {NoStop}%
\bibitem [{\citenamefont {Liska}\ and\ \citenamefont {Gritsev}(2021)}]{Liska2021}%
  \BibitemOpen
  \bibfield  {author} {\bibinfo {author} {\bibfnamefont {D.}~\bibnamefont {Liska}}\ and\ \bibinfo {author} {\bibfnamefont {V.}~\bibnamefont {Gritsev}},\ }\bibfield  {title} {\bibinfo {title} {Hidden symmetries, the bianchi classification and geodesics of the quantum geometric ground-state manifolds},\ }\bibfield  {journal} {\bibinfo  {journal} {SciPost Physics}\ }\textbf {\bibinfo {volume} {10}},\ \href {https://doi.org/10.21468/scipostphys.10.1.020} {10.21468/scipostphys.10.1.020} (\bibinfo {year} {2021})\BibitemShut {NoStop}%
\bibitem [{\citenamefont {Peotta}\ \emph {et~al.}(2023)\citenamefont {Peotta}, \citenamefont {Huhtinen},\ and\ \citenamefont {T{\"o}rm{\"a}}}]{peotta2023quantum}%
  \BibitemOpen
  \bibfield  {author} {\bibinfo {author} {\bibfnamefont {S.}~\bibnamefont {Peotta}}, \bibinfo {author} {\bibfnamefont {K.-E.}\ \bibnamefont {Huhtinen}},\ and\ \bibinfo {author} {\bibfnamefont {P.}~\bibnamefont {T{\"o}rm{\"a}}},\ }\bibfield  {title} {\bibinfo {title} {Quantum geometry in superfluidity and superconductivity},\ }\href@noop {} {\bibfield  {journal} {\bibinfo  {journal} {arXiv preprint arXiv:2308.08248}\ } (\bibinfo {year} {2023})}\BibitemShut {NoStop}%
\bibitem [{\citenamefont {Alexandrov}\ and\ \citenamefont {Gorsky}(2023)}]{alexandrov2023information}%
  \BibitemOpen
  \bibfield  {author} {\bibinfo {author} {\bibfnamefont {A.}~\bibnamefont {Alexandrov}}\ and\ \bibinfo {author} {\bibfnamefont {A.}~\bibnamefont {Gorsky}},\ }\bibfield  {title} {\bibinfo {title} {Information geometry and synchronization phase transition in the kuramoto model},\ }\bibfield  {journal} {\bibinfo  {journal} {Physical Review E}\ }\textbf {\bibinfo {volume} {107}},\ \href {https://doi.org/10.1103/physreve.107.044211} {10.1103/physreve.107.044211} (\bibinfo {year} {2023})\BibitemShut {NoStop}%
\bibitem [{\citenamefont {Pandey}\ \emph {et~al.}(2020)\citenamefont {Pandey}, \citenamefont {Claeys}, \citenamefont {Campbell}, \citenamefont {Polkovnikov},\ and\ \citenamefont {Sels}}]{Pandey}%
  \BibitemOpen
  \bibfield  {author} {\bibinfo {author} {\bibfnamefont {M.}~\bibnamefont {Pandey}}, \bibinfo {author} {\bibfnamefont {P.~W.}\ \bibnamefont {Claeys}}, \bibinfo {author} {\bibfnamefont {D.~K.}\ \bibnamefont {Campbell}}, \bibinfo {author} {\bibfnamefont {A.}~\bibnamefont {Polkovnikov}},\ and\ \bibinfo {author} {\bibfnamefont {D.}~\bibnamefont {Sels}},\ }\bibfield  {title} {\bibinfo {title} {Adiabatic eigenstate deformations as a sensitive probe for quantum chaos},\ }\bibfield  {journal} {\bibinfo  {journal} {Physical Review X}\ }\textbf {\bibinfo {volume} {10}},\ \href {https://doi.org/10.1103/physrevx.10.041017} {10.1103/physrevx.10.041017} (\bibinfo {year} {2020})\BibitemShut {NoStop}%
\bibitem [{\citenamefont {Surace}\ and\ \citenamefont {Motrunich}(2023)}]{Surace_2023}%
  \BibitemOpen
  \bibfield  {author} {\bibinfo {author} {\bibfnamefont {F.~M.}\ \bibnamefont {Surace}}\ and\ \bibinfo {author} {\bibfnamefont {O.}~\bibnamefont {Motrunich}},\ }\bibfield  {title} {\bibinfo {title} {Weak integrability breaking perturbations of integrable models},\ }\bibfield  {journal} {\bibinfo  {journal} {Physical Review Research}\ }\textbf {\bibinfo {volume} {5}},\ \href {https://doi.org/10.1103/physrevresearch.5.043019} {10.1103/physrevresearch.5.043019} (\bibinfo {year} {2023})\BibitemShut {NoStop}%
\bibitem [{\citenamefont {Vanovac}\ \emph {et~al.}(2024)\citenamefont {Vanovac}, \citenamefont {Surace},\ and\ \citenamefont {Motrunich}}]{vanovac2024}%
  \BibitemOpen
  \bibfield  {author} {\bibinfo {author} {\bibfnamefont {S.}~\bibnamefont {Vanovac}}, \bibinfo {author} {\bibfnamefont {F.~M.}\ \bibnamefont {Surace}},\ and\ \bibinfo {author} {\bibfnamefont {O.}~\bibnamefont {Motrunich}},\ }\href {https://arxiv.org/abs/2406.08730} {\bibinfo {title} {Finite-size generators for weak integrability breaking perturbations in the heisenberg chain}} (\bibinfo {year} {2024}),\ \Eprint {https://arxiv.org/abs/2406.08730} {arXiv:2406.08730 [cond-mat.stat-mech]} \BibitemShut {NoStop}%
\bibitem [{\citenamefont {Orlov}\ \emph {et~al.}(2023)\citenamefont {Orlov}, \citenamefont {Tiutiakina}, \citenamefont {Sharipov}, \citenamefont {Petrova}, \citenamefont {Gritsev},\ and\ \citenamefont {Kurlov}}]{orlov2023}%
  \BibitemOpen
  \bibfield  {author} {\bibinfo {author} {\bibfnamefont {P.}~\bibnamefont {Orlov}}, \bibinfo {author} {\bibfnamefont {A.}~\bibnamefont {Tiutiakina}}, \bibinfo {author} {\bibfnamefont {R.}~\bibnamefont {Sharipov}}, \bibinfo {author} {\bibfnamefont {E.}~\bibnamefont {Petrova}}, \bibinfo {author} {\bibfnamefont {V.}~\bibnamefont {Gritsev}},\ and\ \bibinfo {author} {\bibfnamefont {D.~V.}\ \bibnamefont {Kurlov}},\ }\bibfield  {title} {\bibinfo {title} {Adiabatic eigenstate deformations and weak integrability breaking of heisenberg chain},\ }\href {https://doi.org/10.1103/PhysRevB.107.184312} {\bibfield  {journal} {\bibinfo  {journal} {Phys. Rev. B}\ }\textbf {\bibinfo {volume} {107}},\ \bibinfo {pages} {184312} (\bibinfo {year} {2023})}\BibitemShut {NoStop}%
\bibitem [{\citenamefont {Pozsgay}\ \emph {et~al.}(2024)\citenamefont {Pozsgay}, \citenamefont {Sharipov}, \citenamefont {Tiutiakina},\ and\ \citenamefont {Vona}}]{Pozsgay2024}%
  \BibitemOpen
  \bibfield  {author} {\bibinfo {author} {\bibfnamefont {B.}~\bibnamefont {Pozsgay}}, \bibinfo {author} {\bibfnamefont {R.}~\bibnamefont {Sharipov}}, \bibinfo {author} {\bibfnamefont {A.}~\bibnamefont {Tiutiakina}},\ and\ \bibinfo {author} {\bibfnamefont {I.}~\bibnamefont {Vona}},\ }\bibfield  {title} {\bibinfo {title} {Adiabatic gauge potential and integrability breaking with free fermions},\ }\bibfield  {journal} {\bibinfo  {journal} {arXiv preprint arXiv:2402.12979}\ }\href {https://doi.org/10.48550/arXiv.2402.12979} {10.48550/arXiv.2402.12979} (\bibinfo {year} {2024})\BibitemShut {NoStop}%
\bibitem [{\citenamefont {{Sels}}\ and\ \citenamefont {{Polkovnikov}}(2020)}]{Sels}%
  \BibitemOpen
  \bibfield  {author} {\bibinfo {author} {\bibfnamefont {D.}~\bibnamefont {{Sels}}}\ and\ \bibinfo {author} {\bibfnamefont {A.}~\bibnamefont {{Polkovnikov}}},\ }\bibfield  {title} {\bibinfo {title} {{Dynamical obstruction to localization in a disordered spin chain}},\ }\href@noop {} {\bibfield  {journal} {\bibinfo  {journal} {arXiv e-prints}\ ,\ \bibinfo {eid} {arXiv:2009.04501}} (\bibinfo {year} {2020})},\ \Eprint {https://arxiv.org/abs/2009.04501} {arXiv:2009.04501 [quant-ph]} \BibitemShut {NoStop}%
\bibitem [{\citenamefont {Kim}\ and\ \citenamefont {Polkovnikov}(2024)}]{Kim2023}%
  \BibitemOpen
  \bibfield  {author} {\bibinfo {author} {\bibfnamefont {H.}~\bibnamefont {Kim}}\ and\ \bibinfo {author} {\bibfnamefont {A.}~\bibnamefont {Polkovnikov}},\ }\bibfield  {title} {\bibinfo {title} {Integrability as an attractor of adiabatic flows},\ }\href {https://doi.org/10.1103/PhysRevB.109.195162} {\bibfield  {journal} {\bibinfo  {journal} {Phys. Rev. B}\ }\textbf {\bibinfo {volume} {109}},\ \bibinfo {pages} {195162} (\bibinfo {year} {2024})}\BibitemShut {NoStop}%
\bibitem [{\citenamefont {Kravtsov}\ \emph {et~al.}(2015)\citenamefont {Kravtsov}, \citenamefont {Khaymovich}, \citenamefont {Cuevas},\ and\ \citenamefont {Amini}}]{kravtsov2015random}%
  \BibitemOpen
  \bibfield  {author} {\bibinfo {author} {\bibfnamefont {V.~E.}\ \bibnamefont {Kravtsov}}, \bibinfo {author} {\bibfnamefont {I.~M.}\ \bibnamefont {Khaymovich}}, \bibinfo {author} {\bibfnamefont {E.}~\bibnamefont {Cuevas}},\ and\ \bibinfo {author} {\bibfnamefont {M.}~\bibnamefont {Amini}},\ }\bibfield  {title} {\bibinfo {title} {A random matrix model with localization and ergodic transitions},\ }\href {https://doi.org/10.1088/1367-2630/17/12/122002} {\bibfield  {journal} {\bibinfo  {journal} {New Journal of Physics}\ }\textbf {\bibinfo {volume} {17}},\ \bibinfo {pages} {122002} (\bibinfo {year} {2015})}\BibitemShut {NoStop}%
\bibitem [{\citenamefont {Venturelli}\ \emph {et~al.}(2023)\citenamefont {Venturelli}, \citenamefont {Cugliandolo}, \citenamefont {Schehr},\ and\ \citenamefont {Tarzia}}]{venturelli2023replica}%
  \BibitemOpen
  \bibfield  {author} {\bibinfo {author} {\bibfnamefont {D.}~\bibnamefont {Venturelli}}, \bibinfo {author} {\bibfnamefont {L.~F.}\ \bibnamefont {Cugliandolo}}, \bibinfo {author} {\bibfnamefont {G.}~\bibnamefont {Schehr}},\ and\ \bibinfo {author} {\bibfnamefont {M.}~\bibnamefont {Tarzia}},\ }\bibfield  {title} {\bibinfo {title} {Replica approach to the generalized rosenzweig-porter model},\ }\bibfield  {journal} {\bibinfo  {journal} {SciPost Physics}\ }\textbf {\bibinfo {volume} {14}},\ \href {https://doi.org/10.21468/scipostphys.14.5.110} {10.21468/scipostphys.14.5.110} (\bibinfo {year} {2023})\BibitemShut {NoStop}%
\bibitem [{\citenamefont {Bhattacharjee}\ and\ \citenamefont {Nandy}(2024)}]{bhattacharjee2024krylov}%
  \BibitemOpen
  \bibfield  {author} {\bibinfo {author} {\bibfnamefont {B.}~\bibnamefont {Bhattacharjee}}\ and\ \bibinfo {author} {\bibfnamefont {P.}~\bibnamefont {Nandy}},\ }\href@noop {} {\bibinfo {title} {{Krylov fractality and complexity in generic random matrix ensembles}}} (\bibinfo {year} {2024}),\ \Eprint {https://arxiv.org/abs/2407.07399} {arXiv:2407.07399 [quant-ph]} \BibitemShut {NoStop}%
\bibitem [{\citenamefont {Bogomolny}\ and\ \citenamefont {Sieber}(2018)}]{Bogomolny2018eigenvalue}%
  \BibitemOpen
  \bibfield  {author} {\bibinfo {author} {\bibfnamefont {E.}~\bibnamefont {Bogomolny}}\ and\ \bibinfo {author} {\bibfnamefont {M.}~\bibnamefont {Sieber}},\ }\bibfield  {title} {\bibinfo {title} {Eigenfunction distribution for the {Rosenzweig-Porter} model},\ }\href {https://doi.org/10.1103/PhysRevE.98.032139} {\bibfield  {journal} {\bibinfo  {journal} {Phys. Rev. E}\ }\textbf {\bibinfo {volume} {98}},\ \bibinfo {pages} {032139} (\bibinfo {year} {2018})}\BibitemShut {NoStop}%
\bibitem [{\citenamefont {Tang}\ and\ \citenamefont {Khaymovich}(2022)}]{Tang2022nonergodic}%
  \BibitemOpen
  \bibfield  {author} {\bibinfo {author} {\bibfnamefont {W.}~\bibnamefont {Tang}}\ and\ \bibinfo {author} {\bibfnamefont {I.~M.}\ \bibnamefont {Khaymovich}},\ }\bibfield  {title} {\bibinfo {title} {Non-ergodic delocalized phase with {P}oisson level statistics},\ }\href {https://doi.org/10.22331/q-2022-06-09-733} {\bibfield  {journal} {\bibinfo  {journal} {{Quantum}}\ }\textbf {\bibinfo {volume} {6}},\ \bibinfo {pages} {733} (\bibinfo {year} {2022})}\BibitemShut {NoStop}%
\bibitem [{\citenamefont {Khaymovich}\ and\ \citenamefont {Kravtsov}(2021)}]{Khaymovich2021dynamical}%
  \BibitemOpen
  \bibfield  {author} {\bibinfo {author} {\bibfnamefont {I.~M.}\ \bibnamefont {Khaymovich}}\ and\ \bibinfo {author} {\bibfnamefont {V.~E.}\ \bibnamefont {Kravtsov}},\ }\bibfield  {title} {\bibinfo {title} {{Dynamical phases in a ``multifractal'' {Rosenzweig-Porter} model}},\ }\href {https://doi.org/10.21468/SciPostPhys.11.2.045} {\bibfield  {journal} {\bibinfo  {journal} {SciPost Phys.}\ }\textbf {\bibinfo {volume} {11}},\ \bibinfo {pages} {45} (\bibinfo {year} {2021})}\BibitemShut {NoStop}%
\bibitem [{\citenamefont {von Soosten}\ and\ \citenamefont {Warzel}(2018)}]{vonSoosten2017non}%
  \BibitemOpen
  \bibfield  {author} {\bibinfo {author} {\bibfnamefont {P.}~\bibnamefont {von Soosten}}\ and\ \bibinfo {author} {\bibfnamefont {S.}~\bibnamefont {Warzel}},\ }\bibfield  {title} {\bibinfo {title} {Non-ergodic delocalization in the {Rosenzweig}--{Porter} model},\ }\href {https://doi.org/10.1007/s11005-018-1131-7} {\bibfield  {journal} {\bibinfo  {journal} {Letters in Mathematical Physics}\ ,\ \bibinfo {pages} {1}} (\bibinfo {year} {2018})}\BibitemShut {NoStop}%
\bibitem [{\citenamefont {Altshuler}\ and\ \citenamefont {Kravtsov}(2023)}]{altshuler2023random}%
  \BibitemOpen
  \bibfield  {author} {\bibinfo {author} {\bibfnamefont {B.}~\bibnamefont {Altshuler}}\ and\ \bibinfo {author} {\bibfnamefont {V.}~\bibnamefont {Kravtsov}},\ }\bibfield  {title} {\bibinfo {title} {Random cantor sets and mini-bands in local spectrum of quantum systems},\ }\href {https://doi.org/10.1016/j.aop.2023.169300} {\bibfield  {journal} {\bibinfo  {journal} {Annals of Physics}\ }\textbf {\bibinfo {volume} {456}},\ \bibinfo {pages} {169300} (\bibinfo {year} {2023})}\BibitemShut {NoStop}%
\bibitem [{\citenamefont {Sugiura}\ \emph {et~al.}(2021)\citenamefont {Sugiura}, \citenamefont {Claeys}, \citenamefont {Dymarsky},\ and\ \citenamefont {Polkovnikov}}]{Sugiura_2021}%
  \BibitemOpen
  \bibfield  {author} {\bibinfo {author} {\bibfnamefont {S.}~\bibnamefont {Sugiura}}, \bibinfo {author} {\bibfnamefont {P.~W.}\ \bibnamefont {Claeys}}, \bibinfo {author} {\bibfnamefont {A.}~\bibnamefont {Dymarsky}},\ and\ \bibinfo {author} {\bibfnamefont {A.}~\bibnamefont {Polkovnikov}},\ }\bibfield  {title} {\bibinfo {title} {Adiabatic landscape and optimal paths in ergodic systems},\ }\bibfield  {journal} {\bibinfo  {journal} {Physical Review Research}\ }\textbf {\bibinfo {volume} {3}},\ \href {https://doi.org/10.1103/physrevresearch.3.013102} {10.1103/physrevresearch.3.013102} (\bibinfo {year} {2021})\BibitemShut {NoStop}%
\bibitem [{\citenamefont {Larkin}\ and\ \citenamefont {Ovchinnikov}(1969)}]{larkin1969quasiclassical}%
  \BibitemOpen
  \bibfield  {author} {\bibinfo {author} {\bibfnamefont {A.~I.}\ \bibnamefont {Larkin}}\ and\ \bibinfo {author} {\bibfnamefont {Y.~N.}\ \bibnamefont {Ovchinnikov}},\ }\bibfield  {title} {\bibinfo {title} {Quasiclassical method in the theory of superconductivity},\ }\href {https://api.semanticscholar.org/CorpusID:117608877} {\bibfield  {journal} {\bibinfo  {journal} {Journal of Experimental and Theoretical Physics}\ } (\bibinfo {year} {1969})}\BibitemShut {NoStop}%
\bibitem [{\citenamefont {Parker}\ \emph {et~al.}(2019)\citenamefont {Parker}, \citenamefont {Cao}, \citenamefont {Avdoshkin}, \citenamefont {Scaffidi},\ and\ \citenamefont {Altman}}]{parker2019universal}%
  \BibitemOpen
  \bibfield  {author} {\bibinfo {author} {\bibfnamefont {D.~E.}\ \bibnamefont {Parker}}, \bibinfo {author} {\bibfnamefont {X.}~\bibnamefont {Cao}}, \bibinfo {author} {\bibfnamefont {A.}~\bibnamefont {Avdoshkin}}, \bibinfo {author} {\bibfnamefont {T.}~\bibnamefont {Scaffidi}},\ and\ \bibinfo {author} {\bibfnamefont {E.}~\bibnamefont {Altman}},\ }\bibfield  {title} {\bibinfo {title} {A universal operator growth hypothesis},\ }\bibfield  {journal} {\bibinfo  {journal} {Physical Review X}\ }\textbf {\bibinfo {volume} {9}},\ \href {https://doi.org/10.1103/physrevx.9.041017} {10.1103/physrevx.9.041017} (\bibinfo {year} {2019})\BibitemShut {NoStop}%
\bibitem [{\citenamefont {Evers}\ and\ \citenamefont {Mirlin}(2008)}]{evers2008anderson}%
  \BibitemOpen
  \bibfield  {author} {\bibinfo {author} {\bibfnamefont {F.}~\bibnamefont {Evers}}\ and\ \bibinfo {author} {\bibfnamefont {A.~D.}\ \bibnamefont {Mirlin}},\ }\bibfield  {title} {\bibinfo {title} {Anderson transitions},\ }\href {https://doi.org/10.1103/revmodphys.80.1355} {\bibfield  {journal} {\bibinfo  {journal} {Reviews of Modern Physics}\ }\textbf {\bibinfo {volume} {80}},\ \bibinfo {pages} {1355–1417} (\bibinfo {year} {2008})}\BibitemShut {NoStop}%
\bibitem [{\citenamefont {Serbyn}\ \emph {et~al.}(2015)\citenamefont {Serbyn}, \citenamefont {Papić},\ and\ \citenamefont {Abanin}}]{2015}%
  \BibitemOpen
  \bibfield  {author} {\bibinfo {author} {\bibfnamefont {M.}~\bibnamefont {Serbyn}}, \bibinfo {author} {\bibfnamefont {Z.}~\bibnamefont {Papić}},\ and\ \bibinfo {author} {\bibfnamefont {D.~A.}\ \bibnamefont {Abanin}},\ }\bibfield  {title} {\bibinfo {title} {Criterion for many-body localization-delocalization phase transition},\ }\bibfield  {journal} {\bibinfo  {journal} {Physical Review X}\ }\textbf {\bibinfo {volume} {5}},\ \href {https://doi.org/10.1103/physrevx.5.041047} {10.1103/physrevx.5.041047} (\bibinfo {year} {2015})\BibitemShut {NoStop}%
\bibitem [{\citenamefont {Lim}\ \emph {et~al.}(2024)\citenamefont {Lim}, \citenamefont {Matirko}, \citenamefont {Polkovnikov},\ and\ \citenamefont {Flynn}}]{lim2024defining}%
  \BibitemOpen
  \bibfield  {author} {\bibinfo {author} {\bibfnamefont {C.}~\bibnamefont {Lim}}, \bibinfo {author} {\bibfnamefont {K.}~\bibnamefont {Matirko}}, \bibinfo {author} {\bibfnamefont {A.}~\bibnamefont {Polkovnikov}},\ and\ \bibinfo {author} {\bibfnamefont {M.~O.}\ \bibnamefont {Flynn}},\ }\href {https://arxiv.org/abs/2401.01927} {\bibinfo {title} {Defining classical and quantum chaos through adiabatic transformations}} (\bibinfo {year} {2024}),\ \Eprint {https://arxiv.org/abs/2401.01927} {arXiv:2401.01927 [cond-mat.stat-mech]} \BibitemShut {NoStop}%
\bibitem [{\citenamefont {Fine}\ \emph {et~al.}(2014)\citenamefont {Fine}, \citenamefont {Elsayed}, \citenamefont {Kropf},\ and\ \citenamefont {de~Wijn}}]{fine2014absence}%
  \BibitemOpen
  \bibfield  {author} {\bibinfo {author} {\bibfnamefont {B.~V.}\ \bibnamefont {Fine}}, \bibinfo {author} {\bibfnamefont {T.~A.}\ \bibnamefont {Elsayed}}, \bibinfo {author} {\bibfnamefont {C.~M.}\ \bibnamefont {Kropf}},\ and\ \bibinfo {author} {\bibfnamefont {A.~S.}\ \bibnamefont {de~Wijn}},\ }\bibfield  {title} {\bibinfo {title} {Absence of exponential sensitivity to small perturbations in nonintegrable systems of spins 1/2},\ }\bibfield  {journal} {\bibinfo  {journal} {Physical Review E}\ }\textbf {\bibinfo {volume} {89}},\ \href {https://doi.org/10.1103/physreve.89.012923} {10.1103/physreve.89.012923} (\bibinfo {year} {2014})\BibitemShut {NoStop}%
\bibitem [{\citenamefont {Kukuljan}\ \emph {et~al.}(2017)\citenamefont {Kukuljan}, \citenamefont {Grozdanov},\ and\ \citenamefont {Prosen}}]{kukuljan2017weak}%
  \BibitemOpen
  \bibfield  {author} {\bibinfo {author} {\bibfnamefont {I.}~\bibnamefont {Kukuljan}}, \bibinfo {author} {\bibfnamefont {S.}~\bibnamefont {Grozdanov}},\ and\ \bibinfo {author} {\bibfnamefont {T.}~\bibnamefont {Prosen}},\ }\bibfield  {title} {\bibinfo {title} {Weak quantum chaos},\ }\bibfield  {journal} {\bibinfo  {journal} {Physical Review B}\ }\textbf {\bibinfo {volume} {96}},\ \href {https://doi.org/10.1103/physrevb.96.060301} {10.1103/physrevb.96.060301} (\bibinfo {year} {2017})\BibitemShut {NoStop}%
\bibitem [{\citenamefont {Pappalardi}\ \emph {et~al.}(2020)\citenamefont {Pappalardi}, \citenamefont {Polkovnikov},\ and\ \citenamefont {Silva}}]{pappalardi2020quantum}%
  \BibitemOpen
  \bibfield  {author} {\bibinfo {author} {\bibfnamefont {S.}~\bibnamefont {Pappalardi}}, \bibinfo {author} {\bibfnamefont {A.}~\bibnamefont {Polkovnikov}},\ and\ \bibinfo {author} {\bibfnamefont {A.}~\bibnamefont {Silva}},\ }\bibfield  {title} {\bibinfo {title} {Quantum echo dynamics in the sherrington-kirkpatrick model},\ }\bibfield  {journal} {\bibinfo  {journal} {SciPost Physics}\ }\textbf {\bibinfo {volume} {9}},\ \href {https://doi.org/10.21468/scipostphys.9.2.021} {10.21468/scipostphys.9.2.021} (\bibinfo {year} {2020})\BibitemShut {NoStop}%
\bibitem [{\citenamefont {Berry}\ and\ \citenamefont {Shukla}(2020)}]{Berry_2020}%
  \BibitemOpen
  \bibfield  {author} {\bibinfo {author} {\bibfnamefont {M.}~\bibnamefont {Berry}}\ and\ \bibinfo {author} {\bibfnamefont {P.}~\bibnamefont {Shukla}},\ }\bibfield  {title} {\bibinfo {title} {Quantum metric statistics for random-matrix families},\ }\href {https://doi.org/10.1088/1751-8121/ab91d6} {\bibfield  {journal} {\bibinfo  {journal} {Journal of Physics A: Mathematical and Theoretical}\ }\textbf {\bibinfo {volume} {53}},\ \bibinfo {pages} {275202} (\bibinfo {year} {2020})}\BibitemShut {NoStop}%
\bibitem [{\citenamefont {Derrida}(1980)}]{Derrida_1980}%
  \BibitemOpen
  \bibfield  {author} {\bibinfo {author} {\bibfnamefont {B.}~\bibnamefont {Derrida}},\ }\bibfield  {title} {\bibinfo {title} {Random-energy model: Limit of a family of disordered models},\ }\href {https://doi.org/10.1103/PhysRevLett.45.79} {\bibfield  {journal} {\bibinfo  {journal} {Phys. Rev. Lett.}\ }\textbf {\bibinfo {volume} {45}},\ \bibinfo {pages} {79} (\bibinfo {year} {1980})}\BibitemShut {NoStop}%
\bibitem [{\citenamefont {Motamarri}\ \emph {et~al.}(2022)\citenamefont {Motamarri}, \citenamefont {Gorsky},\ and\ \citenamefont {Khaymovich}}]{motamarri2022localization}%
  \BibitemOpen
  \bibfield  {author} {\bibinfo {author} {\bibfnamefont {V.}~\bibnamefont {Motamarri}}, \bibinfo {author} {\bibfnamefont {A.~S.}\ \bibnamefont {Gorsky}},\ and\ \bibinfo {author} {\bibfnamefont {I.}~\bibnamefont {Khaymovich}},\ }\bibfield  {title} {\bibinfo {title} {Localization and fractality in disordered russian doll model},\ }\bibfield  {journal} {\bibinfo  {journal} {SciPost Physics}\ }\textbf {\bibinfo {volume} {13}},\ \href {https://doi.org/10.21468/scipostphys.13.5.117} {10.21468/scipostphys.13.5.117} (\bibinfo {year} {2022})\BibitemShut {NoStop}%
\bibitem [{\citenamefont {Skvortsov}\ \emph {et~al.}(2022)\citenamefont {Skvortsov}, \citenamefont {Amini},\ and\ \citenamefont {Kravtsov}}]{skvortsov2022sensitivity}%
  \BibitemOpen
  \bibfield  {author} {\bibinfo {author} {\bibfnamefont {M.~A.}\ \bibnamefont {Skvortsov}}, \bibinfo {author} {\bibfnamefont {M.}~\bibnamefont {Amini}},\ and\ \bibinfo {author} {\bibfnamefont {V.~E.}\ \bibnamefont {Kravtsov}},\ }\bibfield  {title} {\bibinfo {title} {Sensitivity of (multi)fractal eigenstates to a perturbation of the hamiltonian},\ }\bibfield  {journal} {\bibinfo  {journal} {Physical Review B}\ }\textbf {\bibinfo {volume} {106}},\ \href {https://doi.org/10.1103/physrevb.106.054208} {10.1103/physrevb.106.054208} (\bibinfo {year} {2022})\BibitemShut {NoStop}%
\bibitem [{\citenamefont {Iliesiu}\ \emph {et~al.}(2021)\citenamefont {Iliesiu}, \citenamefont {Mezei},\ and\ \citenamefont {Sárosi}}]{iliesiu2021}%
  \BibitemOpen
  \bibfield  {author} {\bibinfo {author} {\bibfnamefont {L.~V.}\ \bibnamefont {Iliesiu}}, \bibinfo {author} {\bibfnamefont {M.}~\bibnamefont {Mezei}},\ and\ \bibinfo {author} {\bibfnamefont {G.}~\bibnamefont {Sárosi}},\ }\href {https://arxiv.org/abs/2107.06286} {\bibinfo {title} {The volume of the black hole interior at late times}} (\bibinfo {year} {2021}),\ \Eprint {https://arxiv.org/abs/2107.06286} {arXiv:2107.06286 [hep-th]} \BibitemShut {NoStop}%
\bibitem [{\citenamefont {Bhattacharjee}(2023)}]{bhattacharjee2023}%
  \BibitemOpen
  \bibfield  {author} {\bibinfo {author} {\bibfnamefont {B.}~\bibnamefont {Bhattacharjee}},\ }\href {https://arxiv.org/abs/2302.07228} {\bibinfo {title} {A lanczos approach to the adiabatic gauge potential}} (\bibinfo {year} {2023}),\ \Eprint {https://arxiv.org/abs/2302.07228} {arXiv:2302.07228 [quant-ph]} \BibitemShut {NoStop}%
\bibitem [{\citenamefont {Takahashi}\ and\ \citenamefont {del Campo}(2024)}]{Takahashi_2024}%
  \BibitemOpen
  \bibfield  {author} {\bibinfo {author} {\bibfnamefont {K.}~\bibnamefont {Takahashi}}\ and\ \bibinfo {author} {\bibfnamefont {A.}~\bibnamefont {del Campo}},\ }\bibfield  {title} {\bibinfo {title} {Shortcuts to adiabaticity in krylov space},\ }\bibfield  {journal} {\bibinfo  {journal} {Physical Review X}\ }\textbf {\bibinfo {volume} {14}},\ \href {https://doi.org/10.1103/physrevx.14.011032} {10.1103/physrevx.14.011032} (\bibinfo {year} {2024})\BibitemShut {NoStop}%
\bibitem [{\citenamefont {Dumitriu}\ and\ \citenamefont {Edelman}(2002)}]{dumitriu2002matrix}%
  \BibitemOpen
  \bibfield  {author} {\bibinfo {author} {\bibfnamefont {I.}~\bibnamefont {Dumitriu}}\ and\ \bibinfo {author} {\bibfnamefont {A.}~\bibnamefont {Edelman}},\ }\bibfield  {title} {\bibinfo {title} {Matrix models for beta ensembles},\ }\href {https://doi.org/10.1063/1.1507823} {\bibfield  {journal} {\bibinfo  {journal} {Journal of Mathematical Physics}\ }\textbf {\bibinfo {volume} {43}},\ \bibinfo {pages} {5830–5847} (\bibinfo {year} {2002})}\BibitemShut {NoStop}%
\bibitem [{\citenamefont {Balasubramanian}\ \emph {et~al.}(2023)\citenamefont {Balasubramanian}, \citenamefont {Magan},\ and\ \citenamefont {Wu}}]{balasubramanian2023tridiagonalizing}%
  \BibitemOpen
  \bibfield  {author} {\bibinfo {author} {\bibfnamefont {V.}~\bibnamefont {Balasubramanian}}, \bibinfo {author} {\bibfnamefont {J.~M.}\ \bibnamefont {Magan}},\ and\ \bibinfo {author} {\bibfnamefont {Q.}~\bibnamefont {Wu}},\ }\bibfield  {title} {\bibinfo {title} {Tridiagonalizing random matrices},\ }\href {https://doi.org/10.1103/PhysRevD.107.126001} {\bibfield  {journal} {\bibinfo  {journal} {Phys. Rev. D}\ }\textbf {\bibinfo {volume} {107}},\ \bibinfo {pages} {126001} (\bibinfo {year} {2023})}\BibitemShut {NoStop}%
\bibitem [{\citenamefont {Saad}\ \emph {et~al.}(2019)\citenamefont {Saad}, \citenamefont {Shenker},\ and\ \citenamefont {Stanford}}]{saad2019jt}%
  \BibitemOpen
  \bibfield  {author} {\bibinfo {author} {\bibfnamefont {P.}~\bibnamefont {Saad}}, \bibinfo {author} {\bibfnamefont {S.~H.}\ \bibnamefont {Shenker}},\ and\ \bibinfo {author} {\bibfnamefont {D.}~\bibnamefont {Stanford}},\ }\href {https://arxiv.org/abs/1903.11115} {\bibinfo {title} {Jt gravity as a matrix integral}} (\bibinfo {year} {2019}),\ \Eprint {https://arxiv.org/abs/1903.11115} {arXiv:1903.11115 [hep-th]} \BibitemShut {NoStop}%
\bibitem [{\citenamefont {Mehta}(2004)}]{mehta2004random}%
  \BibitemOpen
  \bibfield  {author} {\bibinfo {author} {\bibfnamefont {M.~L.}\ \bibnamefont {Mehta}},\ }\href@noop {} {\emph {\bibinfo {title} {Random matrices}}}\ (\bibinfo  {publisher} {Elsevier},\ \bibinfo {year} {2004})\BibitemShut {NoStop}%
\end{thebibliography}%

\newpage
\onecolumngrid
\appendix

\begin{center}
	{\Large Supplemental Material for ``Hilbert space geometry and quantum chaos''}
\end{center}

\section{Quantum Geometric tensor in terms of matrix elements of the Hamiltonian} \label{App:Ham}

Here we want to show the formulae Eq.(\ref{Geom}). To do so, we start with the original definition:
\begin{equation}\label{eq:matricder}
   g_{\alpha \beta}^{n}(\lambda)=\left< \partial_{\alpha }n|\partial_{\beta }n \right>-\left< \partial_{\alpha }n|n\right> \left< n|\partial_{\beta }n \right>=\sum_m \left(\left< \partial_{\alpha }n| m \right> \left<m | \partial_{\beta }n \right>-\left< \partial_{\alpha }n| m \right> \left<m |n\right> \left< n| m \right> \left<m |\partial_{\beta }n \right>\right),
\end{equation}
where we inserted the identity $\sum_m \left| m \right> \left< m \right|$ in each overlap. Now we can differentiate stationary Schrödinger equation Eq.(\ref{eq:Shr})
\begin{equation}
    \partial_{\alpha} H \left|n \right>+H \left|\partial_{\alpha}  n \right>=\partial_{\alpha} E_n \left|n \right> + E_n \left|\partial_{\alpha} n \right>
\end{equation}
calculating an overlap with another eigenvector $\left|m \right>$, we can get
\begin{eqnarray}
    \left<m | \partial_{\alpha }n \right>=\frac{\left<m \right| \partial_{\alpha} H \left| n \right>}{E_n-E_m},~~~n \neq m.
\end{eqnarray}
Inserting this expression into eq.(\ref{eq:matricder}) we get the desired formula
\begin{eqnarray}
    G_{\alpha \beta}(\lambda)=\frac{1}{N} \sum_{m \neq n} \frac{\left<n \right| \partial_{\alpha} H \left| m\right>\left<m \right| \partial_{\beta} H \left| n \right>}{(E_n-E_m)^2}.
\end{eqnarray}

Generally this expression needs regularization if there are degeneracies in the spectrum~\cite{Sugiura_2021} (see also Appendix~\ref{App:mu}).

\section{Calculation of the 2d QGT components} \label{calcul}
In this appendix, we provide detailed derivation of the components of quantum geometric tensor as a function of parameter $r$. We also show that due to symmetry of the problem, the components are angle independent.

Firstly, we change the basis of the random matrices in the following way:
\begin{equation}\label{rot}
 \tilde{H}_{x}=\cos\phi H_x+\sin \phi H_y,~\tilde{H}_{y}= \cos \phi H_y-\sin\phi H_x,
\end{equation}then the integration measure changes as
\begin{equation}
 \mathcal{D} H_x~ \mathcal{D} H_y= J~ \mathcal{D} \tilde{H}_x~ \mathcal{D} \tilde{H}_y.
\end{equation}
 Here we notice that the Jacobian $J$ does not depend on $\tilde{H}_x$ and $\tilde{H}_{y}$, since the change of coordinates is linear. Therefore, the overall Jacobian factor, does not affect the averages over matrix ensembles.
\subsection{\texorpdfstring{$\overline{G}_{r\phi}$}{TEXT} component}
Let's start with the mixed component.  In terms of new variables, $r, \phi$ the $\bar{g}_{r\phi}$ component can be written in the following way:
\begin{equation}
\overline{G}_{r\phi} = Z^{-1}  \frac{r}{N} \sum_{m\neq n}\int  \frac{\langle n| \tilde{H}_x|m\rangle \langle m| \tilde{H}_y|n\rangle}{(H_0+r \tilde{H}_x)_{nm}^2} \rho(H_0,\tilde{H}_x,\tilde{H}_y) \mathcal{D} H_0  \mathcal{D} H_x \mathcal{D} H_y 
\end{equation}
Notice that $\rho(H_0,H_x, H_y)=\rho(H_0,\tilde{H}_x,\tilde{H}_y)$ , since $Tr(H_x^2+H_y^2)=Tr(\tilde{H}_x^2+\tilde{H}_y^2)$. Thus, the average is
\begin{equation}
\overline{G}_{r\phi} = Z^{-1}  \frac{r}{N} \sum\limits_{m \neq n}\int  \frac{\langle n| \tilde{H}_x|m\rangle \langle m| \tilde{H}_y|n\rangle}{(H_0+r \tilde{H}_x)_{nm}^2} e^{-\frac{N}{2}~Tr\left( H_0^2+\tilde{H}_x^2 +\tilde{H_y}^2 \right) } \mathcal{D} H_0  \mathcal{D} \tilde{H}_x \mathcal{D} \tilde{H}_y 
\end{equation}

In the pre-exponential factor, only the second term in the numerator depends on the matrix $\tilde{H}_y$.  So integrating over $\tilde{H}_{y}$ matrix we obtain:
\begin{equation}
\int \mathcal{D} \tilde{H}_{y} \langle n| \tilde{H}_y |m\rangle e^{-\frac{N}{2} Tr(\tilde{H}_y^2)}=0
\end{equation}Finally we obtain  $\bar{g}_{r\phi}=0$.
\subsection{\texorpdfstring{$\overline{G}_{rr}$}{TEXT} component}\label{App:grr}

For  the $\bar{g}_{rr} $ component we have:
\begin{equation}
\overline{G}_{rr} = Z^{-1} \frac{1}{N} \sum_{m\neq n}\int  \frac{\langle n| \tilde{H}_x|m\rangle \langle m| \tilde{H}_x|n\rangle}{(H_0+r \tilde{H}_x)_{nm}^2} \rho(H_0,\tilde{H}_x,\tilde{H}_y) \mathcal{D} H_0  \mathcal{D} H_x \mathcal{D} H_y 
\end{equation}

The integration over $\tilde{H}_y$ returns unity. Then, the rest of the integral reads as:
\begin{equation}
\overline{G}_{rr} = Z^{-1}\frac{1}{N}  \sum_{m\neq n}\int  \frac{\langle n| \tilde{H}_x|m\rangle \langle m| \tilde{H}_x|n\rangle}{(H_0+r \tilde{H}_x)_{nm}^2} e^{-\frac{N}{2}~Tr\left(H_0^2+\tilde{H}_x^2 \right) } \mathcal{D} H_0  \mathcal{D} H_x 
\end{equation}
It is convenient to change the variables of integration again:
\begin{equation}
H_0 \rightarrow H =H_0+r \tilde{H}_x~~~~
\end{equation}
Because this transformation is linear there is no Jacobian and hence:
\begin{equation}
\overline{G}_{rr} = Z^{-1}\frac{1}{N}  \sum_{m\neq n}\int  \frac{\langle n| \tilde{H}_x|m\rangle \langle m| \tilde{H}_x|n\rangle}{(H)_{nm}^2} e^{-\frac{N}{2}~Tr\left((H-r\tilde{H}_x)^2+\tilde{H}_x^2 \right) } \mathcal{D} H  \mathcal{D} H_x .
\end{equation}
Upon some simple calculations the metric has simple Gaussian form in the coordinates $\tilde H$ and $H$:
\begin{equation}
\overline{G}_{rr} = Z^{-1}  \frac{1}{N(r^2+1)}\sum_{m\neq n}\int  \frac{\langle n| \tilde{H}|m\rangle \langle m| \tilde{H}|n\rangle}{(H)_{nm}^2} e^{-\frac{N}{2}~Tr\left(\frac{1}{r^2+1} H^2 +\tilde{H}^2 \right) } \mathcal{D} H  \mathcal{D} \tilde{H}.
\end{equation}
Now we integrate over $\tilde{H}$, the only  $\tilde{H} $  dependent part of the integral  is
\begin{equation}\label{avv}
\frac{\int \langle n| \tilde{H}|m\rangle \langle m| \tilde{H}|n\rangle e^{-\frac{N}{2} ~Tr\tilde{H}^2} \mathcal{D}\tilde{H}}{\int  e^{-\frac{1}{2}N ~Tr\tilde{H}^2} \mathcal{D}\tilde{H}}.
\end{equation}
Moreover we can explicitly write the components of eigenvectors as
\begin{equation}\label{eq:matr_el}
 \langle n| \tilde{H}|m\rangle \langle m| \tilde{H}|n\rangle=(n^{*}_i H_{ij} m_j)~(m^{*}_k H_{kl} n_l),
\end{equation}
using correlator of GUE matrices $\left< H_{ij}H_{kl} \right>_{GUE} =\frac{1}{N}\delta_{il} \delta_{jk}$, the average of Eq.\eqref{eq:matr_el} can be written in the following form:

\begin{gather}\label{2point}
\left< (n^{*}_i H_{ij} m_j)~(m^{*}_k H_{kl} n_l) \right>_{GUE}=\frac{1}{N}(n^{*}_i  m_j)~(m^{*}_k n_l)\delta_{il} \delta_{jk}=\frac{1}{N}\bra{n} \ket{n} \bra{m} \ket{m}=\frac{1}{N}.
\end{gather}
So, the metric is rewritten as
\begin{equation}
\overline{G}_{rr} = Z^{-1}  \frac{1}{N^2(r^2+1)}\sum_{m\neq n}\int  \frac{1}{(E_n-E_m)^2} e^{-\frac{N}{2(r^2+1)} ~Tr\left(H^2  \right) } \mathcal{D} H .
\end{equation}
In order to proceed further, let's rewrite the expression by the eigenvalue integration:
\begin{equation}
\overline{G}_{rr} = \frac{1}{N^2(r^2+1)}~\sum_{m\neq n} \frac{\int  \frac{1}{(E_n-E_m)^2} \prod\limits_{i>j} (E_i-E_j)^2 e^{-\frac{N}{2(r^2+1)} ~ \sum\limits_{i=1}^N E_i^2   }  \prod\limits_{i} dE_i }{\int \prod\limits_{i>j} (E_i-E_j)^2 e^{-\frac{N}{2(r^2+1)} ~ \sum\limits_{i=1}^N E_i^2   }  \prod\limits_{i} dE_i},
\end{equation}
where $\prod\limits_{i>j} (E_i-E_j)^2$ is a standard Vandermonde determinant. Now changing the variables by $E_{i}=E'_i \sqrt{\frac{r^2+1}{N}}$, we obtain 
\begin{equation}
\begin{split}
&\overline{G}_{rr} = \frac{1}{N^2(r^2+1)^2}~\sum_{m\neq n} \frac{\int  \frac{1}{(E'_n-E'_m)^2} \prod\limits_{i>j} (E'_i-E'_j)^2 e^{-\frac{1}{2} ~ \sum\limits_{i=1}^N E_i^{'~2}   }  \prod\limits_{i} dE'_i }{\int \prod\limits_{i>j} (E'_i-E'_j)^2 e^{-\frac{1}{2} ~ \sum\limits_{i=1}^N E_i^{'~2}   }  \prod\limits_{i} dE'_i}= \\
&=\frac{1}{N^2(r^2+1)^2} \sum_{m\neq n} \left< \frac{1}{(\lambda_n-\lambda_m)^2}\right>_{GUE}=\frac{N-1}{2(r^2+1)^2}\;,
\end{split}
\end{equation}
where this average can be calculated using the ‘virial relation’ (see e.g. \cite{mehta2004random}):
\begin{equation}\label{virial}
 \sum\limits_{n=1}^{N} \sum\limits_{m\neq n} \left< \frac{1}{(E_n-E_m)^2} \right>_{GUE} =\frac{N^2(N-1)}{2}.
\end{equation}
It depends on the level $n$ and the size of the matrix $N$. Thus, we found the explicit expression for metric up to constant to be found numerically.

\subsection{ \texorpdfstring{$\overline{G}_{\phi\phi}$}{TEXT} component}
The calculation of the angle-angle component can be performed in the similar manner, following steps of the previous Appendix \ref{App:grr}
The $\bar{g}_{\phi \phi} $ reads:
\begin{equation}
\overline{G}_{\phi \phi} = Z^{-1}\frac{1}{N} r^2\sum_{m\neq n}\int  \frac{\langle n| \tilde{H}_y|m\rangle \langle m| \tilde{H}_y|n\rangle}{(H_0+r \tilde{H}_x)_{nm}^2} \rho(H_0,\tilde{H}_x,\tilde{H}_y) \mathcal{D} H_0  \mathcal{D} \tilde{H}_x \mathcal{D} \tilde{H}_y.
\end{equation}
Using the same average over $\tilde{H}_y$ as in Eq. (\ref{2point}) we obtain:
\begin{equation}
\overline{G}_{\phi \phi} =  Z^{-1} \frac{r^2}{N^2}  \sum_{m\neq n}\int  \frac{1}{(H_0+r \tilde{H}_x)_{nm}^2} e^{-\frac{N}{2}Tr\left( H_0^2+\tilde{H}_x^2\right)}\mathcal{D} H_0  \mathcal{D} \tilde{H}_x ,
\end{equation}
again changing the coordinates in the following way
\begin{equation}
H_0 \rightarrow H=H_0+r \tilde{H}_x.
\end{equation}
Thus, we will change the integration  $\mathcal{D} H_0  \mathcal{D} \tilde{H}_x  \rightarrow \mathcal{D} H  \mathcal{D} \tilde{H}_x $ and using the same technique as in the previous subsection we obtain

\begin{equation}
\overline{G}_{\phi \phi} = r^2 \frac{N-1}{2(r^2+1)} \;,
\end{equation}
with the same constant we defined above.

\section{\texorpdfstring{$N=2$}{TEXT} Integrability breaking metric}\label{App:metrn2}
In this appendix, we present the derivation of the metric tensor for the random matrix system with the integrable point. Similar to the purely chaotic case, we perform an averaging over random matrices. As a result, we identify an isomorphic two-dimensional manifold. Additionally, we observe a conical singularity at the integrable point on this manifold.
\subsection{\texorpdfstring{$\overline{G}_{\phi\phi}$}{TEXT} component}

To calculate, $\overline{G}_{\phi \phi}$ we will use the expression derived in the previous appendix:
\begin{equation}
\overline{G}_{\phi \phi} =  Z^{-1} \frac{1}{N} r^2  \sum_{m\neq n}\int  \frac{1}{(H_0+r \tilde{H}_x)_{nm}^2} e^{-\frac{N}{2}Tr\left( H_0^2+2\tilde{H}_x^2\right)}\mathcal{D} H_0  \mathcal{D} \tilde{H}_x \;,
\end{equation}
 where now $\mathcal{D} H_0=\prod\limits_{i} d h_i$.
In $N=2$ case, the matrix $H_0$ corresponds to the integrable point, that can be parametrised by two independent Gaussian distributed parameters. While perturbation that is represented by GUE matrix of the size $2$ is parametrised by $4$ parameters:
\begin{equation}
H_0=\begin{pmatrix}
h_1&&0\\
0&&h_2
\end{pmatrix}~~~~\tilde{H}_x=\begin{pmatrix}
a&&c e^{i \phi}\\
c e^{-i \phi }&&b
\end{pmatrix}\;.
\end{equation}
Then the integration measure simply reads as
\begin{equation}
\mathcal{D} H_0  \mathcal{D} \tilde{H}_x =d h_1~d h_2~da ~db~c dc~d\phi \;.
\end{equation}
Notice that eigenvalues of the Hamiltonian do not depend on the parameter $\phi$, so we can trivially integrate over $\phi$. After we change the variables
$$
h_1+h_2=x_1~~,~~h_1-h_2=x_2\;,
$$
$$
a+b=y_1~~,~~a-b=y_2\;.
$$
Thus, in the new variables the integral reads as
\begin{equation}
\overline{G}_{\phi \phi} = \frac{1}{2}Z^{-1}r^2\int\frac{e^{-\frac{1}{4}\left(x_1^2+x_2^2 \right)-\frac{1}{2}\left(y_1^2+y_2^2+4 c^2 \right)}}{(x_2+r y_2)^2+4 c^2 r^2 }~ c dc~ dx_2~ dy_2~ dx_1~ dy_1\;.
\end{equation}We can integrate over $x_1$ and $y_1$  and change the variables again :
$$
x_2+r y_2=z_1~~,~~x_2-r y_2=z_2.
$$

After the integration, we arrive at the final expression
\begin{equation}\label{64}
\overline{G}_{\phi \phi}=r \frac{1}{2~\sqrt{2} } arctg\left( \frac{\sqrt{2}}{r}\right).
\end{equation}

\subsection{\texorpdfstring{$\overline{G}_{rr}$}{TEXT} component}
The consideration of $rr$ component can be done in the similar manner. We start with the expression:  
\begin{equation}
\overline{G}_{r r} = Z^{-1}  \sum_{m\neq n}\int  \frac{\langle n| \tilde{H}_x|m\rangle \langle m| \tilde{H}_x|n\rangle}{(H_0+r \tilde{H}_x)_{nm}^2} e^{-\frac{1}{2}~Tr\left(H_0^2+2\tilde{H}_x^2 \right) } \mathcal{D} H_0  \mathcal{D} H_x \;.
\end{equation}Using parametrisation we discussed for the $\phi \phi $ component of the metric and the same change of variables we obtain the expression:
\begin{equation}
\overline{G}_{r r} = Z^{-1}  \int\frac{e^{-\frac{1}{4}\left(x_1^2+x_2^2 \right)-\frac{1}{2}\left(y_1^2+y_2^2+4 c^2 \right)}~~x_2^2 c^2}{\left((x_2+r y_2)^2+4 c^2 r^2\right)^{2}}~ c  dc~ dx_2~ dy_2~ dx_1~ dy_1.
\end{equation}
Calculating the Gaussian integrals we arrive at
\begin{equation}\label{71}
\overline{G}^{(a)}_{r r}=\frac{1}{4} \left(\frac{arcctg\left(\frac{r}{\sqrt{2}}\right)}{\sqrt{2}r}-\frac{1}{2+r^2} \right).
\end{equation}

\subsection{Embedding in 3d pseudo-Euclidean space}
Following the procedure we discussed in the main text we can find the radius component of the embedding: 
\begin{equation}
R^2(r)=r \frac{1}{2 \sqrt{2}} \arctan\left( \frac{\sqrt{2}}{r}\right),
\end{equation} and the form of the surface $Z(r)$ can be found from Eq.(\ref{Emb}), which reduces to a differential equation:
\begin{equation}
\left(\frac{dZ}{dr}\right)^2=\frac{1}{8}\left[ \frac{ ~\cot^{-1}\left( \frac{r}{\sqrt{2}}\right) }{\sqrt{2} r}-\frac{2\sqrt{2} r\left(2+r^2\right)^{-2}}{~\cot^{-1}\left( \frac{r}{\sqrt{2}}\right)}\right].
\end{equation}
Although we can not calculate the integral exactly,  we investigate the asymptotic limits. Here we consider two opposite limits: at infinite value of the parameter $r$ and when $r$ goes to zero.
In the infinite limit  $r \rightarrow \infty $ 
\begin{equation}
\begin{split}
&\overline{G}_{rr}(r) \rightarrow \frac{1}{4 r^4}~,~~\frac{\overline{G}_{\phi \phi}(r)}{r^2} \rightarrow \frac{1}{2 r^2},\\
&\Rightarrow R(\infty)\rightarrow \frac{\sqrt{2}}{2} ~~~\text{and}~~Z(\infty)\rightarrow 1.097,
\end{split}
\end{equation}and in the  $r\rightarrow 0$  limit the functions are the following,
\begin{equation}
\begin{split}
&\overline{G}_{rr}(r) \rightarrow \frac{\pi}{8\sqrt{2} r}-1/4~,~~\frac{\overline{G}_{\phi \phi}(r)}{r^2} \rightarrow \frac{\pi}{4 \sqrt{2} r}-1/4\\
&R(r)\rightarrow\sqrt{\frac{\pi}{4 \sqrt{2}}}\sqrt{r}~,~~Z(r)~\rightarrow\sqrt{\frac{\pi}{4 \sqrt{2}}}\sqrt{r}. 
\end{split}
\end{equation}
These approximate values were obtained by the expansions up to the order $\mathcal{O}(r^{3/2})$.
Therefore, near the point $r=0$ or $Z=R=0$ the angle of slope of this surface of revolution is
\begin{equation}
\frac{dZ}{dR}\left(0\right)=1.
\end{equation}

\section{{Conical singularly}}\label{Conical}

The Gauss-Bonnet theorem allows finding the Euler characteristic $\chi$ of a Riemannian manifold: 
\begin{equation}
\int_M K d A+ \int_{\partial M} k d s = 2 \pi \chi(M),
\end{equation}
where K is the Gaussian curvature of a compact two-dimensional Riemannian manifold M and $ k$ is the geodesic curvature of the boundary $\partial M$.
There is an extension of this theorem that allows to find the Euler characteristic of a manifold with conical singularities:
\begin{equation}
\int_M K d A+ \int_{\partial M} k d s+\sum_i \theta_i = 2 \pi \chi(M),
\end{equation}where $\theta_i$ is an angular defect of the manifold at point i.
For instance, let's consider a two-dimensional space with topology of the cone:
\begin{equation}
d s^2=d \rho^2+\alpha^2 \rho^2 d \phi^2
\end{equation}where $\phi \in [0,2 \pi]$,  $\alpha \in (0,1]$ and $\rho \in [0,1]$. If we make a replacement, $\tilde{\phi}=\alpha \phi$ we will get the flat metric in polar coordinates with $\tilde{\phi}\in [0,2 \pi \alpha]$ . Thus, the angular defect is $\theta=2\pi(1-\alpha)$.

 The embedding of the conical defect
into the 3d pseudo-Euclidean space is described by the equation 
\begin{equation}
    z^2 -\frac{|1-\alpha^2|}{\alpha^2}(x^2+y^2)=0, \qquad z\geq 0
\end{equation}

Using the simple geometry, 
we can find the connection between the angle of the cone and the angular defect
\begin{equation}
\sin{\beta}=\alpha.
\end{equation}As a result, 
\begin{equation}
\theta=2 \pi(1-\sin{\beta}),
\end{equation}
which is an additional contribution to the Euler characteristic of the manifold from the defect.

\section{Quantum geometric tensor with energy cutoff and transition timescales}\label{App:mu}

\begin{figure*}[h!]
  \centering
  \begin{subfigure}[t]{0.45\linewidth}
    \includegraphics[trim=0 0 0 0, clip, width=\columnwidth]{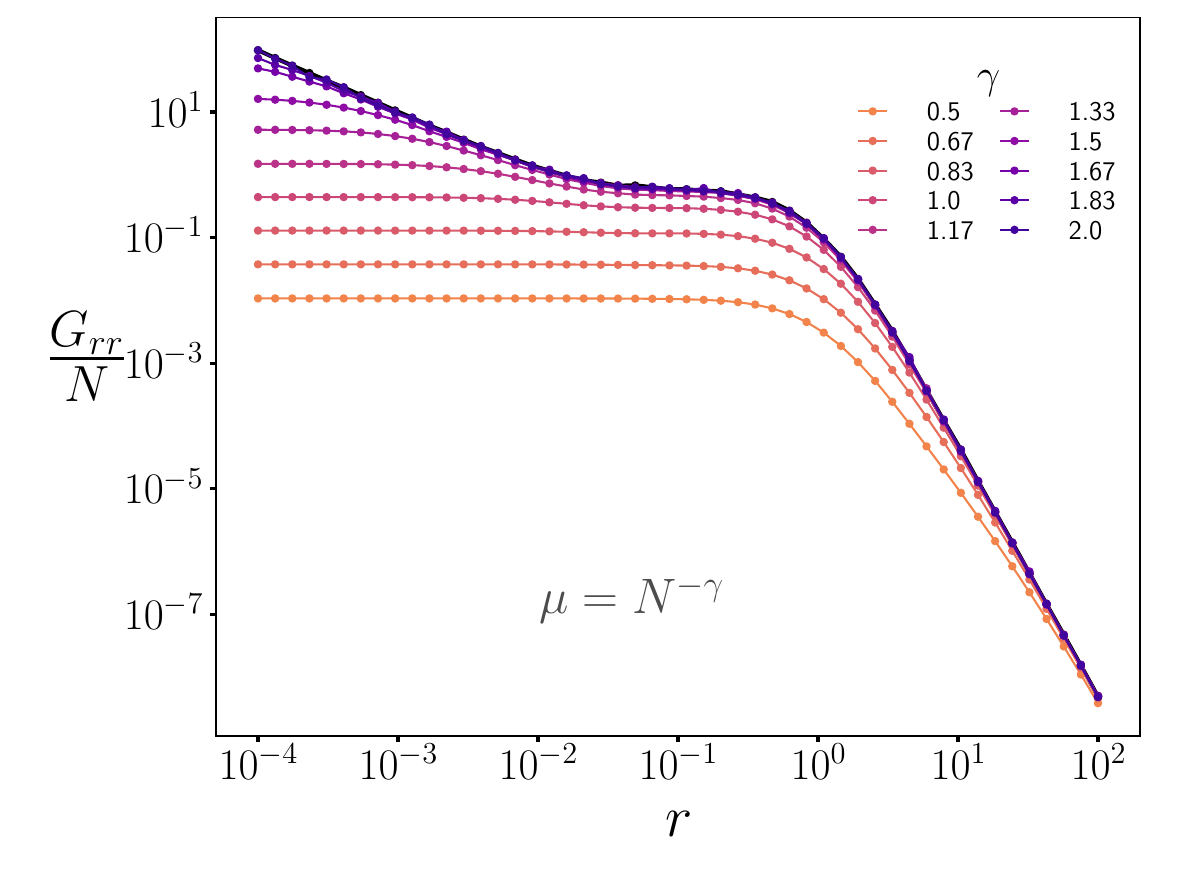}
  \end{subfigure}
  \begin{subfigure}[t]{0.45\linewidth}
    \includegraphics[trim=0 0 0 0, clip, width=\columnwidth]{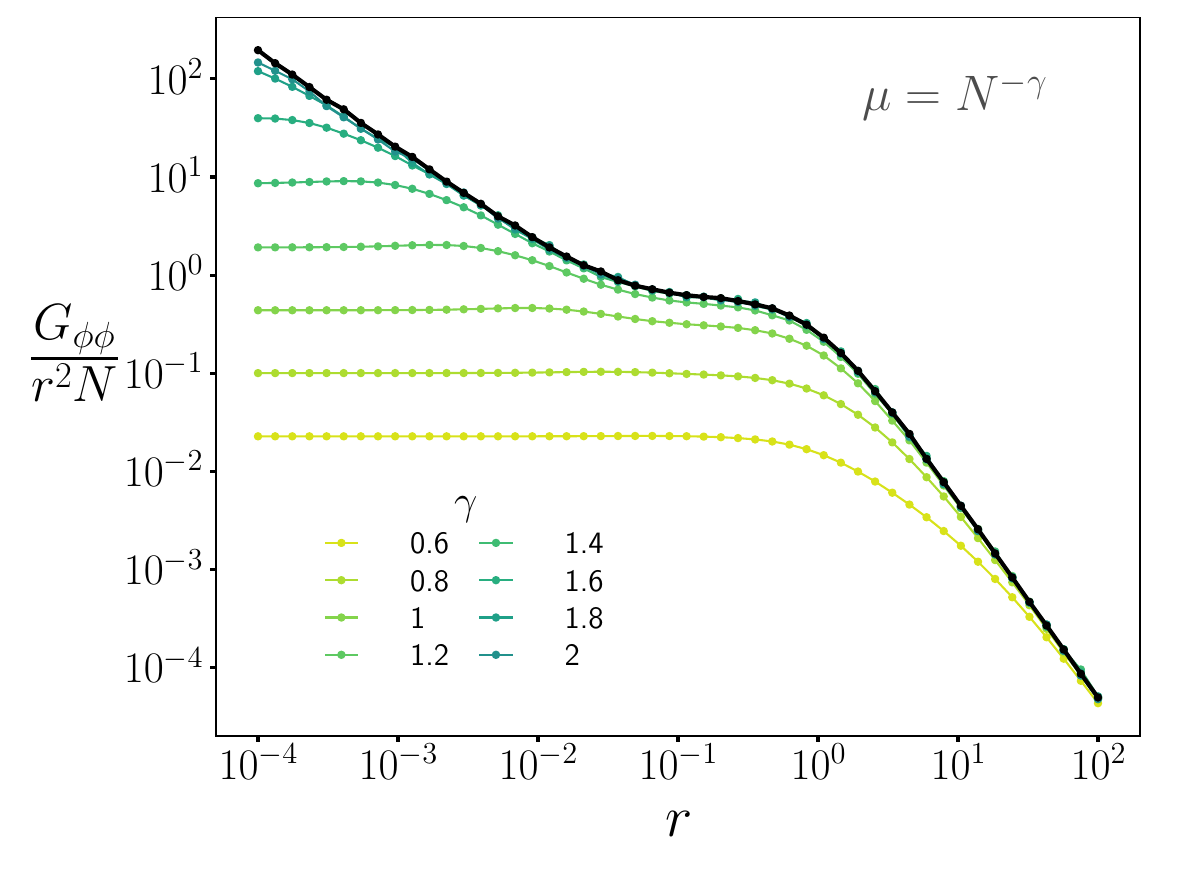}
  \end{subfigure}
  \caption{The $\mu$ dependence of the metric components for different cutoffs with $N=1600$. The black lines corresponds to $\mu=0$.}
  \label{large}
\end{figure*}

To probe the typical timescales for transitions between regimes we introduce the $\mu$ regularization of the metric  having in mind that $t\propto \mu^{-1}$,
\begin{equation}
    g_{\alpha \beta}^{(n)}=\sum_{m\neq n} \frac{\left<n \right| \partial_{\alpha}H \left| m \right>\left<m \right| \partial_{\beta}H \left| n \right>(E_n-E_m)^2}{((E_n-E_m)^2+\mu^2))^2}\;,
\end{equation}
and analyse the behaviour of $g_{rr}(r,\mu), g_{\phi \phi}(r,\mu)$.  
The $\mu$ -dependencies of the metric components at late times are presented at the Fig.\ref{large}. 
We observe that the transition to the integrable regime occurs at $t>N$.

In the parametrization  $\mu \propto N^{-\gamma}$ at small values of $\gamma$ the metric manifests purely chaotic pattern  
and at $\beta \geq 1$ the $1/r$ behaviour is observable  Fig.\ref{large}. In parametrization $t\propto N^{\gamma}$ the observed  timescales are consistent with known 
scaling for the Thouless time $t_{Th}\propto N^{1-D}$ in the RP model \cite{kravtsov2015random}, where $D$ is the fractal dimension in the corresponding phase. 
Indeed, in the localized(integrable) phase $D=0$ hence the expected transition timescale is $t\propto N$ while in the ergodic delocalized phase $D=1$ hence the corresponding timescale is $N^0$.

\end{document}